\documentclass[twocolumn]{aastex61}\usepackage[]{graphicx}\usepackage[]{color}
\makeatletter
\def\maxwidth{ %
  \ifdim\Gin@nat@width>\linewidth
    \linewidth
  \else
    \Gin@nat@width
  \fi
}
\makeatother

\definecolor{fgcolor}{rgb}{0.345, 0.345, 0.345}

\usepackage{framed}
\makeatletter
 {\par\unskip\endMakeFramed%
 \at@end@of@kframe}
\makeatother

\definecolor{shadecolor}{rgb}{.97, .97, .97}
\definecolor{messagecolor}{rgb}{0, 0, 0}
\definecolor{warningcolor}{rgb}{1, 0, 1}
\definecolor{errorcolor}{rgb}{1, 0, 0}

\usepackage{alltt}
\usepackage{amsmath,amsfonts,amssymb}
\usepackage{xspace}
\usepackage{longtable}
\usepackage{txfonts}

\newcommand\nsvs{NSVS 14256825\xspace}
\def\tv#1{{\pmb #1}}
\def\idm#1{\mbox{\scriptsize #1}}
\newcommand\corr[1]{{ #1}}

\received{December 7, 2016}
\revised{January 6, 2017}
\accepted{\today}
\submitjournal{ApJ}

\shorttitle{\nsvs and circumbinary planet}
\shortauthors{Nasiroglu et al.}

\IfFileExists{upquote.sty}{\usepackage{upquote}}{}
\begin{document}


\title{Is there a circumbinary planet around NSVS~14256825?}



\correspondingauthor{Aga S\l{}owikowska}
\email{aga@astro.ia.uz.zgora.pl}

\author{Ilham Nasiroglu}
\affil{Departments of Astronomy and Astrophysics,
Ataturk University
Yakutiye, 25240, Erzurum, Turkey}

\author{Krzysztof Go\'zdziewski}
\affil{Centre for Astronomy, Faculty of Physics, Astronomy and Applied Informatics,
N. Copernicus University, 
Grudziadzka 5, 87-100 Toru{\'n}, Poland}

\author{Aga S\l{}owikowska, Krzysztof Krzeszowski, Micha\l{} \.Zejmo}
\affil{Janusz Gil Institute of Astronomy,
University of Zielona G{\'o}ra, Prof. Szafrana 2,
65-516 Zielona G{\'o}ra, Poland}

\author{Staszek Zola}
\affil{Astronomical Observatory,
Jagiellonian University, Orla 171, 
30-244 Krak{\'o}w, Poland}

\author{Huseyin ER}
\affil{Departments of Astronomy and Astrophysics,
Ataturk University
Yakutiye, 25240, Erzurum, Turkey}

\author{Waldemar Og\l{}oza, Marek Dr{\'o}\.zd\.z}
\affil{Mt. Suhora Observatory,
Pedegogical University, Podchorazych 2, 
30-084 Krak{\'o}w, Poland}

\author{Dorota Koziel-Wierzbowska, Bartlomiej Debski}
\affil{Astronomical Observatory,
Jagiellonian University, ul. Orla 171,
30-244 Krak{\'o}w, Poland}

\author{Nazli Karaman}
\affil{Physics Department,
Adiyaman University
Merkez, 02040, Adiyaman, Turkey}



\begin{abstract}
The cyclic behaviour of (O--C) residuals of eclipse timings in the sdB+M eclipsing binary \nsvs{} was previously attributed to one or two Jovian-type circumbinary planets.
We report 83 new eclipse timings that not only fill in the gaps in those already published but also extend the time span of the (O--C) diagram by three years.  Based on the archival and our new data spanning over more than 17 years we re-examined the up to date system (O--C). The data revealed systematic, quasi-sinusoidal variation deviating from an older linear ephemeris by about 100~s. It also exhibits  a maximum in the (O-C) near JD 2,456,400 that was previously unknown. \corr{We consider two most credible explanations of the (O-C) variability: the light propagation time due to the presence of an invisible companion in a distant circumbinary orbit, and  magnetic cycles reshaping one of the binary components, known as the Applegate or Lanza--Rodon\'o effect. We found that the latter mechanism is unlikely due to the insufficient energy budget of the M-dwarf secondary.  In the framework of the third-body hypothesis, we obtained meaningful constraints on the Keplerian parameters of a putative companion and its mass.}
Our best-fitting model indicates that the observed quasi-periodic (O-C) variability can be explained by the presence of a brown dwarf with the minimal mass of 15 Jupiter masses rather than a planet, orbiting the binary in a moderately elliptical orbit ($e\simeq $0.175) with the period of $\sim$ 10 years. \corr{Our analysis rules out two planets model proposed earlier.}
\end{abstract}




\section{Introduction}
\label{sec:introduction}

The number of discovered exoplanets around binary systems increases rapidly. These discoveries have sparked a rising interest on this subject among  researchers, and consequently, they drive the development of new detection techniques. Studies of circumbinary planets (CBPs) have taken us much closer toward answering the fundamental questions how such planets form and evolve.  The properties of circumbinary planets are likely different than these orbiting isolated stars \citep{Lee2009}.

The most remarkable discovery made with the transit method with the \textit{Kepler} satellite is the discovery of a circumbinary planet transiting across both stars of the close binary system Kepler-16 (AB) \citep{Doyle2011}. The transits in this system leave no doubt about the existence of planets in the so-called ``P-type'' orbits, i.e. circumbinary orbits. So far the longest-period transiting CBP is Kepler-1647 with the orbital period of $\sim1100$ days \citep{Kostov2016a}. 

Even before the \textit{Kepler} discoveries, timing observations have provided evidence of planets orbiting binary systems. Hints for such companions were reported by \cite{Deeg2008} and \cite {Lee2009} for the eclipsing binaries CM Dra and HW Vir, respectively. The presence of these objects are indicated through the Light Travel Time (LTT) effect indicated by the variations in the timings of eclipse minima w.r.t. the linear ephemeris (O-C). Quasi-periodic variations of the (O-C) can result from the gravitational tug due to distant planets (companions), which leads to swinging of the eclipsing binary, and causing the eclipses to appear slightly earlier or later w.r.t the linear ephemeris. The LTT effect can be measured with a high accuracy and used to infer the presence of planetary-mass companions around binary stars \citep{Irwin1952, Horner2012, Gozdziewski2012, Gozdziewski2015}. In contrast to other techniques, the timing method is sensitive to massive extrasolar planets in long-period orbits. Futhermore, for low--mass binaries the amplitude of the LTT effect increases \citep{Ribas2005,Pribulla2012}.

\corr{Recently, a number of} planetary mass companions orbiting the cataclysmic variables (CVs) and post-common envelope binaries (PCEBs) have been reported. For instance, two planets for NN Ser \citep{Beuermann2010} and UZ For \citep{Potter2011}, a single planet for DP Leo \citep{Beuermann2011} and V470 Cam \citep{Beuermann2012} have been claimed. 
A long-term stable system of three planets
hosted by HU Aqr, with the middle one being on a retrograde orbit, was recently proposed by \citet{Gozdziewski2015}.

\nsvs was discovered through the Northern Sky Variability Survey (NSVS) \citep{Wozniak2004}. \citet{Wils2007} identified this system as an eclipsing binary with an amplitude of variations in the range of 13.22--14.03~(V). These authors also presented the first B, V, I$_{\rm{c}}$ light curves and physical parameters of this binary (P=0.110374230(2) days), along with a few eclipse times. \nsvs is a member of the HW Vir family (PCEB) consisting of a OB sub--dwarf and a M dwarf companion (sdOB+dM) \citep{Almeida2012}. The following physical and geometrical parameters were obtained: i = 82.$^{\circ}$5 $\pm$ 0.$^{\circ}$3 (inclination of the system),  M$_1$ = 0.419 $\pm$ 0.070 M$_{\odot}$, R$_1$ = 0.188 $\pm$ 0.010 R$_{\odot}$, M$_2$ = 0.109 $\pm$ 0.023 M$_{\odot}$ R$_2$ = 0.162 $\pm$ 0.008 R$_{\odot}$ (the masses and radii of the components), a = 0.80 $\pm$ 0.04 R$_{\odot}$ (separation between the components) from the photometric and spectroscopic observations \citep{Almeida2012}. 

The eclipse times of \nsvs have been reported by \cite{Wils2007, Kilkenny2012, Beuermann2012, Almeida2013} and \cite{Lohr2014}. \cite{Qian2010} and \cite{Zhu2011} also argued  for a O--C cyclic variation, however they have not published any supporting data yet. \cite{Kilkenny2012}, reported an increasing orbital period  of this system with a rate of $\sim~1.1~\times~10^{-10}~\rm{ss}^{-1}$. \cite{Beuermann2012} detected cyclic period changes and suggested the presence of a single circumbinary planet of $\sim 12~\rm{M_{Jup}}$ with a period of $\sim 20$~yrs. \cite{Almeida2013} presented a few additional eclipse times and argued for the presence of two circumbinary planets with periods of $\sim 3.5$~yrs and $\sim 6.7$~yrs, and masses of $\sim 3~\rm{M_{Jup}}$ and $\sim 8.0~\rm{M_{Jup}}$, respectively. \cite{Wittenmyer2013} presented a dynamical analysis of the orbital stability of the two planet model proposed by \cite{Almeida2013}. They found that this model is extremely unstable on time scale of less than a thousand years. Moreover, \cite{Hinse2014} also performed a detailed data analysis of the timing measurements of this system. They concluded that the time span of eclipse time variations is not long enough neither to explain any particular one--planet model nor provide a convincing evidence for a second planetary companion. Recently, \cite{Lohr2014} presented many new eclipse times of \nsvs from the SuperWASP archive. Their measurements obtained between 2006 and 2011 confirm the overall trend already seen in the O--C diagram.

In this study, we present 83 new mid-eclipse times of \nsvs obtained between 2009-08-21 and 2016-11-03 that together with the literature data give 153 eclipses over the time span of 17 years. We combined our new data with the previously published measurements to analyse the orbital period variations of this system. In Section \ref{sec:newphot} we present the observations and data reduction process together with the methodology used to obtain the eclipse times. Section \ref{sec:model} presents the procedure applied to examine the period variations, while the results are gathered in Section \ref{sec:results}. In Sections \ref{sec:discussion} and \ref{sec:conclusions} we discuss and conclude our findings. We include some additional materials in the on-line Appendix.

\section{New Photometry of \nsvs}
\label{sec:newphot}

We performed photometric 
observations of \nsvs between 2009-08-21 and 2016-11-03 with five different telescopes: the 1.3~m telescope at the Skinakas Observatory (SKO, Creete, Greece), the 0.5~m telescope at the Astronomical Observatory of the Jagiellonian University (KRK, Krak\'ow, Poland), the 0.6~m telescope at the Mt. Suhora Observatory (SUH, Koninki, Poland), the 0.6~m telescope at the Adiyaman University Observatory (ADYU60, Adiyaman, Turkey) and with the 1~m telescope at the TUBITAK National Observatory (TUG, Antalya, Turkey).

Between 2009 and 2013, observations were performed using the SKO, KRK and SUH telescopes, while these taken between 2014 and 2016, with the ADYU60 and TUG telescopes. We gathered  data  with the following CCD cameras: the Andor iKon DZ-936B-BV (KRK), the Andor DZ436 (SKO), the Apoge Alta U47 (SUH), the Andor iKon-M934 (ADYU60) and the SI1100 (TUG).  A summary of observations is given in Tab.~\ref{tab:observations}, where the start observing date, the cycle number, eclipse type (primary -- 1, secondary -- 2), filter used, exposure time and readout time are listed. 

The CCD data were reduced with the pipeline developed using Python, IRAF and Sextractor software. The usual bias and dark subtraction as well as flat-field correction were applied to all images. For Andor CCDs dark counts were negligible and therefore, only bias subtraction was done.  
A nearby constant star in the field of view, comparable to the target star in brightness and color, was chosen as the comparison star. 
Since our major goal was to obtain differential photometry only, we did not observe any photometric standard stars. For each night a light curve was constructed consisting of extracted magnitude differences and time in the form of JD UTC. The mid-exposure times were taken.

\begin{figure*}[!h]
    \centering
	\includegraphics[scale=0.76]{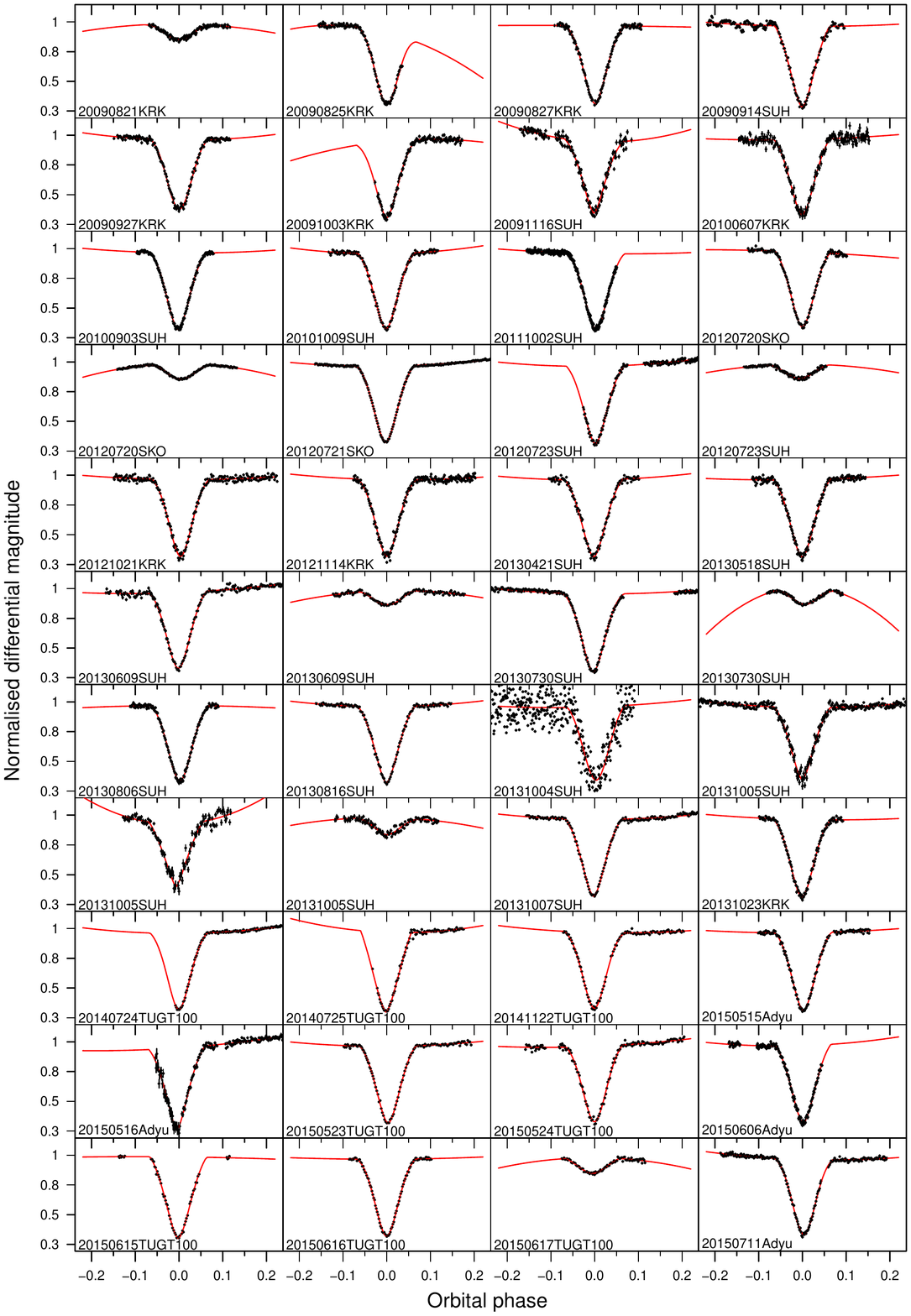}
    \caption{Primary and secondary eclipses of \nsvs from the 1-m telescope of the TUBITAK National Observatory (TUG), the 0.6-m telescope of the Adiyaman University Observatory (ADYU60), the 1.3-m telescope of the Skinakas Observatory (SKO), the 0.5-m telescope of the Astronomical Observatory of the Jagiellonian University (KRK) and the 0.6-m telescope of the Mt. Suhora Observatory (SUH), fitted with the equation as described in Section 2. of \cite{Beuermann2012}.}
    \label{fig:lc_1}
\end{figure*}

\begin{figure*}
  \centering
	\includegraphics[scale=0.76]{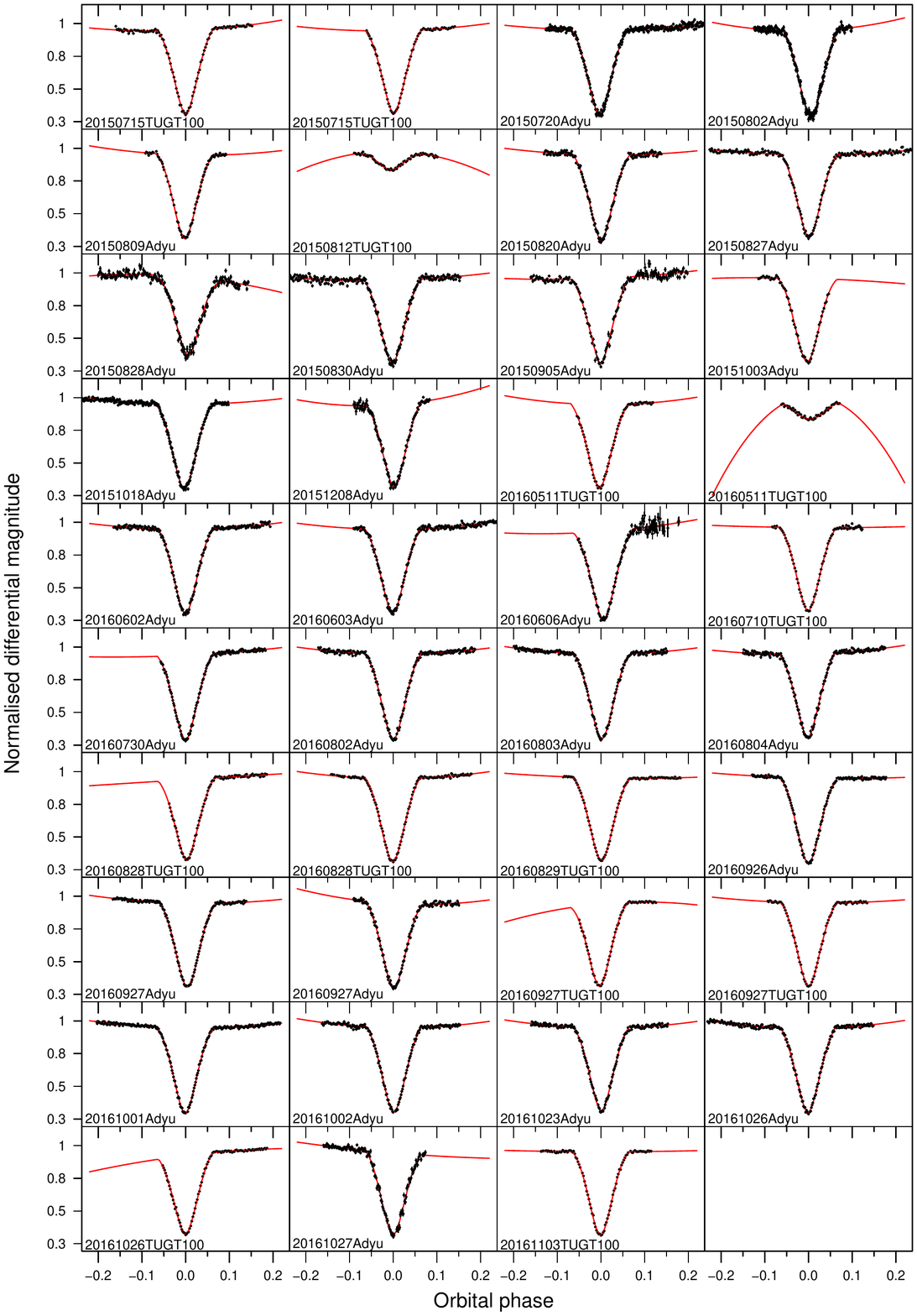}
    \caption{Primary and secondary eclipses of \nsvs from the 1-m telescope of the TUBITAK National Observatory (TUG), the 0.6-m telescope of the Adiyaman University Observatory (ADYU60), the 1.3-m telescope of the Skinakas Observatory (SKO), the 0.5-m telescope of the Astronomical Observatory of the Jagiellonian University (KRK) and the 0.6-m telescope of the Mt. Suhora Observatory (SUH), fitted with the equation as described in Section 2. of \cite{Beuermann2012}.}
    \label{fig:lc_2}
\end{figure*}

We  modelled the shapes of the eclipses with a modified and truncated inverted Gaussian $G(\tau)$ multiplied by a polynomial, as described in Section 2. of \cite{Beuermann2012}. \corr{The model involves 8 parameters, including eclipse minimum time ($T_{\rm{obs}}$) denoted as $p_1$ in \cite{Beuermann2012}. The resulting parameters values and their respective uncertainties (including $\sigma_{T_{\rm{obs}}}$) were obtained from the fitting procedure as a term of the resulting covariance matrix from the nonlinear least squares fitting algorithm}.
Figs.~\ref{fig:lc_1} and \ref{fig:lc_2} show the observed light curves and model fits.
The derived eclipse times were converted to the barycentric dynamical time (BJD), using the FK5 sky coordinates of \nsvs ($\alpha = 20^{\rm h}20^{\rm m}00^{\rm s}.458$,  $\delta = +04^{\circ}37^{'}56^{''}\!.50$) and the geodetic coordinates of each given observatory, with the help of the numerical procedure developed by \cite{Eastman2010}.  The eclipse minimum times, together with their respective errors, obtained from our new measurements are listed in Tab.~\ref{tab:eclipses}, while all timings (including these published in the literature) are gathered in Tab.~\ref{tab:eclipses_all} (available in the on-line version only). The cycle numbers in  Tabs.~\ref{tab:eclipses} and \ref{tab:eclipses_all} are given according to the ephemeris from \cite{Beuermann2012}. \corr{We would like to note that we have three pairs of simultaneous observations
of \nsvs with the TUG and Adiyaman telescopes, i.e. these for cycle numbers: 30669, 30670 and 30931. For each pair the difference between derived $T_{\rm obs}$ agrees  within their errors derived from the fit, i.e. less than two seconds.}

\begin{table}
\centering
\caption{List of new \nsvs eclipse times. Cycle number, time of the minimum,
 its error, type of the eclipse (1 -- primary, 2 -- secondary) and references are
 given.  References are for the observatories: (5) the Astronomical
 Obs. of the Jagiellonian Univ., (6) the Mt.  Suhora
 Obs., (7) the Skinakas Obs., (8) the TUBITAK National
 Obs. and (9) the Adiyaman Univ. Obs.
} 
\label{tab:eclipses}
\begin{tabular}{rcccc}
  \hline
Cycle & BJD & Error [days] & Eclipse Type & Ref \\ 
  \hline
7167.5 & 2455065.315208 & 0.000082 & 2 & 5 \\ 
  7204.0 & 2455069.343870 & 0.000036 & 1 & 5 \\ 
  7223.0 & 2455071.440996 & 0.000014 & 1 & 5 \\ 
  7386.0 & 2455089.432002 & 0.000024 & 1 & 6 \\ 
  7503.0 & 2455102.345843 & 0.000018 & 1 & 5 \\ 
  7557.0 & 2455108.306027 & 0.000041 & 1 & 5 \\ 
  7955.0 & 2455152.234856 & 0.000018 & 1 & 6 \\ 
  9797.0 & 2455355.544034 & 0.000035 & 1 & 5 \\ 
  10593.0 & 2455443.401924 & 0.000010 & 1 & 6 \\ 
  10918.0 & 2455479.273564 & 0.000012 & 1 & 6 \\ 
  14162.0 & 2455837.327516 & 0.000016 & 1 & 6 \\ 
  16808.0 & 2456129.377577 & 0.000015 & 1 & 7 \\ 
  16808.5 & 2456129.432822 & 0.000023 & 2 & 7 \\ 
  16817.0 & 2456130.370981 & 0.000006 & 1 & 7 \\ 
  16835.0 & 2456132.357723 & 0.000017 & 1 & 6 \\ 
  16835.5 & 2456132.412907 & 0.000033 & 2 & 6 \\ 
  17650.0 & 2456222.312685 & 0.000020 & 1 & 5 \\ 
  17867.0 & 2456246.263846 & 0.000021 & 1 & 5 \\ 
  19301.0 & 2456404.540320 & 0.000024 & 1 & 6 \\ 
  19545.0 & 2456431.471682 & 0.000019 & 1 & 6 \\ 
  19744.0 & 2456453.436075 & 0.000040 & 1 & 6 \\ 
  19744.5 & 2456453.491275 & 0.000099 & 2 & 6 \\ 
  20206.0 & 2456504.428991 & 0.000015 & 1 & 6 \\ 
  20206.5 & 2456504.484030 & 0.000044 & 2 & 6 \\ 
  20269.0 & 2456511.382498 & 0.000014 & 1 & 6 \\ 
  20360.0 & 2456521.426538 & 0.000013 & 1 & 6 \\ 
  20803.0 & 2456570.322299 & 0.000098 & 1 & 6 \\ 
  20812.0 & 2456571.315608 & 0.000025 & 1 & 6 \\ 
  20812.5 & 2456571.370781 & 0.000113 & 2 & 6 \\ 
  20813.0 & 2456571.425973 & 0.000046 & 1 & 6 \\ 
  20830.0 & 2456573.302321 & 0.000014 & 1 & 6 \\ 
  20975.0 & 2456589.306568 & 0.000016 & 1 & 5 \\ 
  23458.0 & 2456863.365367 & 0.000033 & 1 & 8 \\ 
  23467.0 & 2456864.358634 & 0.000032 & 1 & 8 \\ 
  24553.0 & 2456984.224844 & 0.000017 & 1 & 8 \\ 
  26132.0 & 2457158.505405 & 0.000012 & 1 & 9 \\ 
  26141.0 & 2457159.498659 & 0.000030 & 1 & 9 \\ 
  26205.0 & 2457166.562708 & 0.000012 & 1 & 8 \\ 
  26213.0 & 2457167.445649 & 0.000021 & 1 & 8 \\ 
   \hline
\end{tabular}
\end{table}

\addtocounter{table}{-1}

\begin{table}
\centering
\caption{continued...} 
\begin{tabular}{rcccc}
\hline
Cycle & BJD & Error [days] & Eclipse Type & Ref \\ 
  \hline
  26331.0 & 2457180.469794 & 0.000026 & 1 & 9 \\ 
  26413.0 & 2457189.520546 & 0.000020 & 1 & 8 \\ 
  26422.0 & 2457190.513847 & 0.000010 & 1 & 8 \\ 
  26422.5 & 2457190.569067 & 0.000075 & 2 & 8 \\ 
  26648.0 & 2457215.458344 & 0.000012 & 1 & 9 \\ 
  26683.0 & 2457219.321420 & 0.000012 & 1 & 8 \\ 
  26684.0 & 2457219.431837 & 0.000013 & 1 & 8 \\ 
  26730.0 & 2457224.509018 & 0.000014 & 1 & 9 \\ 
  26846.0 & 2457237.312397 & 0.000017 & 1 & 9 \\ 
  26911.0 & 2457244.486749 & 0.000012 & 1 & 9 \\ 
  26937.5 & 2457247.411658 & 0.000056 & 2 & 8 \\ 
  27009.0 & 2457255.303376 & 0.000015 & 1 & 9 \\ 
  27073.0 & 2457262.367301 & 0.000016 & 1 & 9 \\ 
  27082.0 & 2457263.360683 & 0.000034 & 1 & 9 \\ 
  27099.0 & 2457265.237010 & 0.000020 & 1 & 9 \\ 
  27154.0 & 2457271.307578 & 0.000031 & 1 & 9 \\ 
  27408.0 & 2457299.342580 & 0.000028 & 1 & 9 \\ 
  27543.0 & 2457314.243090 & 0.000010 & 1 & 9 \\ 
  28004.0 & 2457365.125453 & 0.000022 & 1 & 9 \\ 
  29411.5 & 2457520.476902 & 0.000061 & 2 & 8 \\ 
  29412.0 & 2457520.532048 & 0.000015 & 1 & 8 \\ 
  29611.0 & 2457542.496481 & 0.000012 & 1 & 9 \\ 
  29620.0 & 2457543.489845 & 0.000014 & 1 & 9 \\ 
  29647.0 & 2457546.469948 & 0.000039 & 1 & 9 \\ 
  29956.0 & 2457580.575537 & 0.000013 & 1 & 8 \\ 
  30135.0 & 2457600.332481 & 0.000014 & 1 & 9 \\ 
  30163.0 & 2457603.422952 & 0.000012 & 1 & 9 \\ 
  30172.0 & 2457604.416311 & 0.000011 & 1 & 9 \\ 
  30180.0 & 2457605.299328 & 0.000013 & 1 & 9 \\ 
  30397.0 & 2457629.250499 & 0.000009 & 1 & 8 \\ 
  30399.0 & 2457629.471256 & 0.000020 & 1 & 8 \\ 
  30408.0 & 2457630.464585 & 0.000006 & 1 & 8 \\ 
  30660.0 & 2457658.278862 & 0.000008 & 1 & 9 \\ 
  30669.0 & 2457659.272248 & 0.000008 & 1 & 9 \\ 
  30669.0 & 2457659.272263 & 0.000014 & 1 & 8 \\ 
  30670.0 & 2457659.382587 & 0.000009 & 1 & 8 \\ 
  30670.0 & 2457659.382609 & 0.000015 & 1 & 9 \\ 
  30705.0 & 2457663.245692 & 0.000008 & 1 & 9 \\ 
  30714.0 & 2457664.239053 & 0.000010 & 1 & 9 \\ 
  30904.0 & 2457685.210139 & 0.000012 & 1 & 9 \\ 
  30931.0 & 2457688.190200 & 0.000015 & 1 & 8 \\ 
  30931.0 & 2457688.190212 & 0.000011 & 1 & 9 \\ 
  30941.0 & 2457689.293987 & 0.000027 & 1 & 9 \\ 
  31004.0 & 2457696.247504 & 0.000009 & 1 & 8 \\ 
   \hline
\end{tabular}
\end{table}

\section{LTT models of the (O--C)}
\label{sec:model}

To model the eclipses ephemeris with the presence of a hypothetical third body we used the following formulae
\begin{equation}
T_{\idm{eph}}(L) = t_0 + P_{\idm{bin}} L + \gamma_{\rm tb}(L) , 
\label{eq:eph}
\end{equation}
where $\gamma_{\rm tb}$ represents the Light Travel Time (LTT) term \citep{Irwin1952}. In our formulation this term is parametrised by the Keplerian orbital elements of the third body companion in orbit around the mass center of the binary \citep{Gozdziewski2012}:
\[
 \gamma_{\rm tb}(t) = K [ \sin\omega ( \cos E(t) -e ) +
  \cos\omega \sqrt{1-e^2} \sin E(t) ],
\]
where $K$ is the semi-amplitude of the LTT signal, $e$, $\omega$, $P$, $\tau$ are the eccentricity, periastron argument, orbital period and the time of periastron passage of the {\em relative} orbit of the putative companion w.r.t. the binary. 
We note that $P$ and $\tau$ are introduced indirectly through the Kepler equation
\[
 \frac{2\pi}{P}(t-\tau) = E - e \sin E.
\]

Due to very different time scales of orbital motion, the binary is represented as a point with  the total mass of both stellar components equal to 0.528 M$_\Sun$ \citep{Almeida2012}. Furthermore, to account for small eccentricity, we introduce  Poincar\'e elements
$ (x \equiv e \cos\omega, y \equiv e\sin\omega)$
which make it possible to get rid of weakly constrained eccentricity and pericenter argument $\omega$ for quasi-circular and moderately eccentric orbits.

To express the (O--C) variability through $\gamma_{\rm pl}(L(t))$, we optimize the likelihood function $\cal L$,
\def\tv#1{{\pmb #1}}
\begin{equation}
\log {\cal L}({\cal D}|\tv{\xi}) =   
-\frac{1}{2} \chi^2
     -\frac{1}{2}\sum_{i}{N} \log {\sigma_{i}^2}
     -\frac{1}{2} N \log{2\pi},
\label{eq:Lfun}
\end{equation}
where the $\chi^2$ function
\begin{equation}
\chi^2({\cal D},{\tv{\xi}}) = \sum_i^N\frac{{[\mbox{O}({\cal D})-\mbox{C}(\tv{\xi})]}_{i}^2}{\sigma_i^2}
\label{eq:chi2}
\end{equation}
depends on model parameters through $(\mbox{O--C})_{i} \equiv (T_{\rm obs}(L_i) - T_{\rm eph}(L_i)$) and the measurements uncertainties $\sigma_i$, where $i=1\ldots,N$ \citep{Gozdziewski2015}. Here, $\mbox{(O--C)}_{i}$ denotes the deviation of the observed $i$-th eclipse time-mark from its BJD ephemeris (Eq.~\ref{eq:eph}) for cycle $L_i\equiv L(t_i)$. The model parameters vector $\tv{\xi} \equiv (K,P,e,\omega,\tau,P_{bin},t_0,\sigma_f$), and $N$ denotes the number of measurements encoded as data set ${\cal D}$. We note that all parameters of $T_{\rm eph}$ are optimized. This more general form of ${\cal L}$ makes it also
possible to determine the free parameter $\sigma_f$ that scales the raw uncertainties $\sigma_{i}$ in quadrature, such that $\sigma_{i,t}^2 \rightarrow \sigma_{i}^2+\sigma_f^2$ results in $\chi^2_{\nu} \equiv \chi^2/(N-\mbox{dim}\,\tv{\xi}) \sim 1$. 

Optimisation of the dynamical model relies on investigating the space of $8$ free model parameters $\tv{\xi}$, through sampling the posterior probability distribution ${\cal
P}(\tv{\xi}|{\cal D})$ of the parameters $\tv{\xi}$, given the data set ${\cal
D}$: 
$
   {\cal P}(\tv{\xi}|{\cal D}) 
  \propto {\cal P}(\tv{\xi}) {\cal P}({\cal D}|\tv{\xi}),
$
where ${\cal P}(\tv{\xi})$ is the prior, and the sampling data distribution ${\cal P}({\cal D}|\tv{\xi}) \equiv \log{\cal L}({\cal D}|\tv{\xi})$. For all of these parameters, priors have been set as uniform (or uniform improper) through imposing parameters ranges available for the exploration, i.e., $K,P,\tau>0$~d,
 $\sigma_f>0$~d, and $x,y\in [-0.71,0.71]$, $P_{\idm{\rm bin}} \in [0.110, 0.112]$~d, and $\Delta t_0 \in [-0.1,0.1]$~d which is for  the displacement w.r.t. the cycle $L=0$ for the epoch of $T_0=\mbox{BJD}\,2455793.840061$.

We sampled the posterior with the Markov Chain Monte Carlo (MCMC) emcee package of the affine-invariant ensemble sampler \citep{Goodman2010}, kindly provided by \cite{Foreman2013}. 

\section{LTT model results}
\label{sec:results}

\begin{figure}
\includegraphics[angle=0, width=0.47\textwidth]{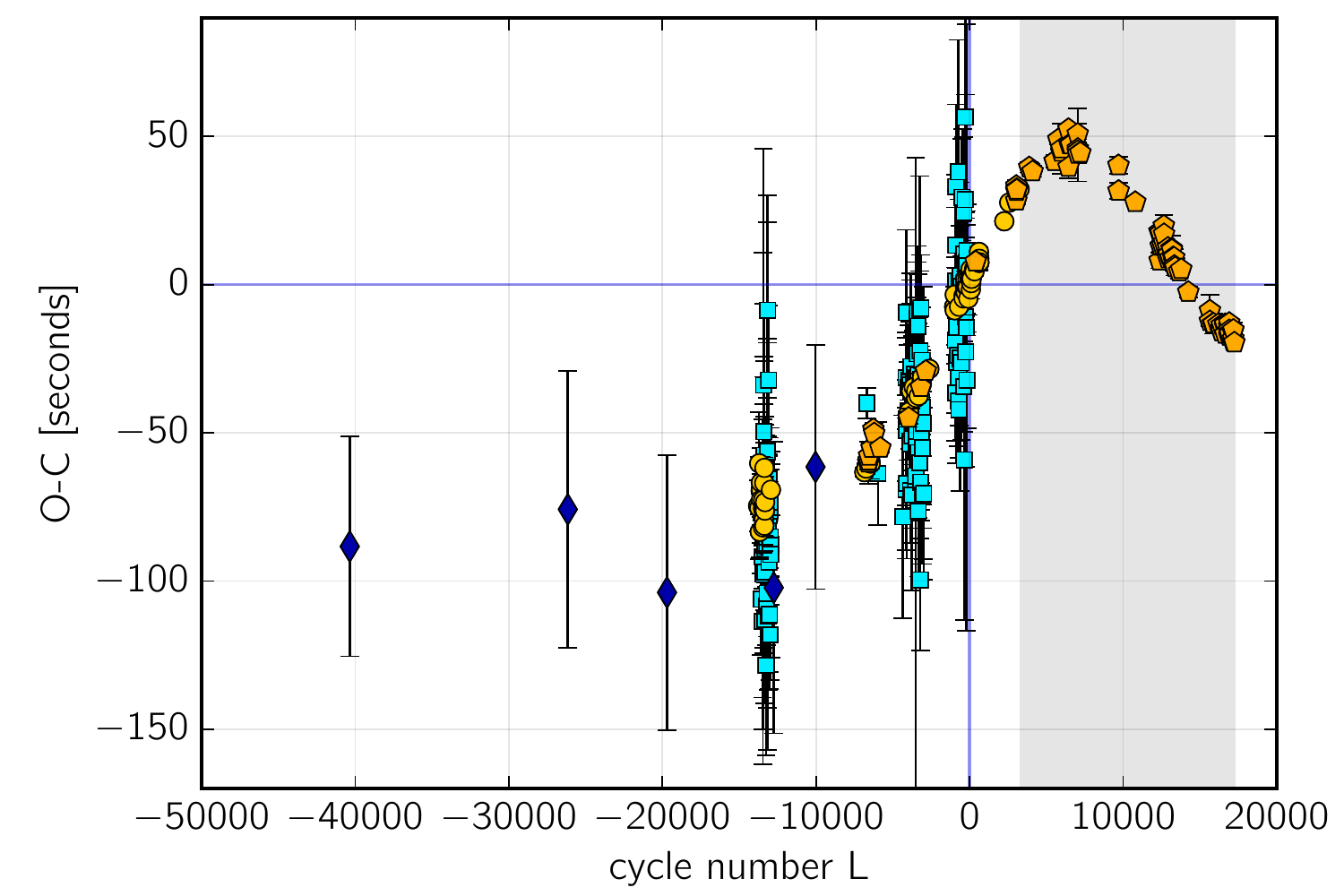}
    \caption{The (O--C) diagram w.r.t. the linear ephemeris for Dataset~A, with all data available in the literature. Filled rectangles are for raw (unbinned) SuperWASP data, filled circles and diamonds are for other measurements prior to epoch of August 2012, and pentagons are for the new timing data in this paper 
    (Tab.~\ref{tab:eclipses}). See the text for details. }
    \label{fig:nsvs_WASP}
\end{figure}

\begin{figure}
\includegraphics[angle=0, width=0.47\textwidth]{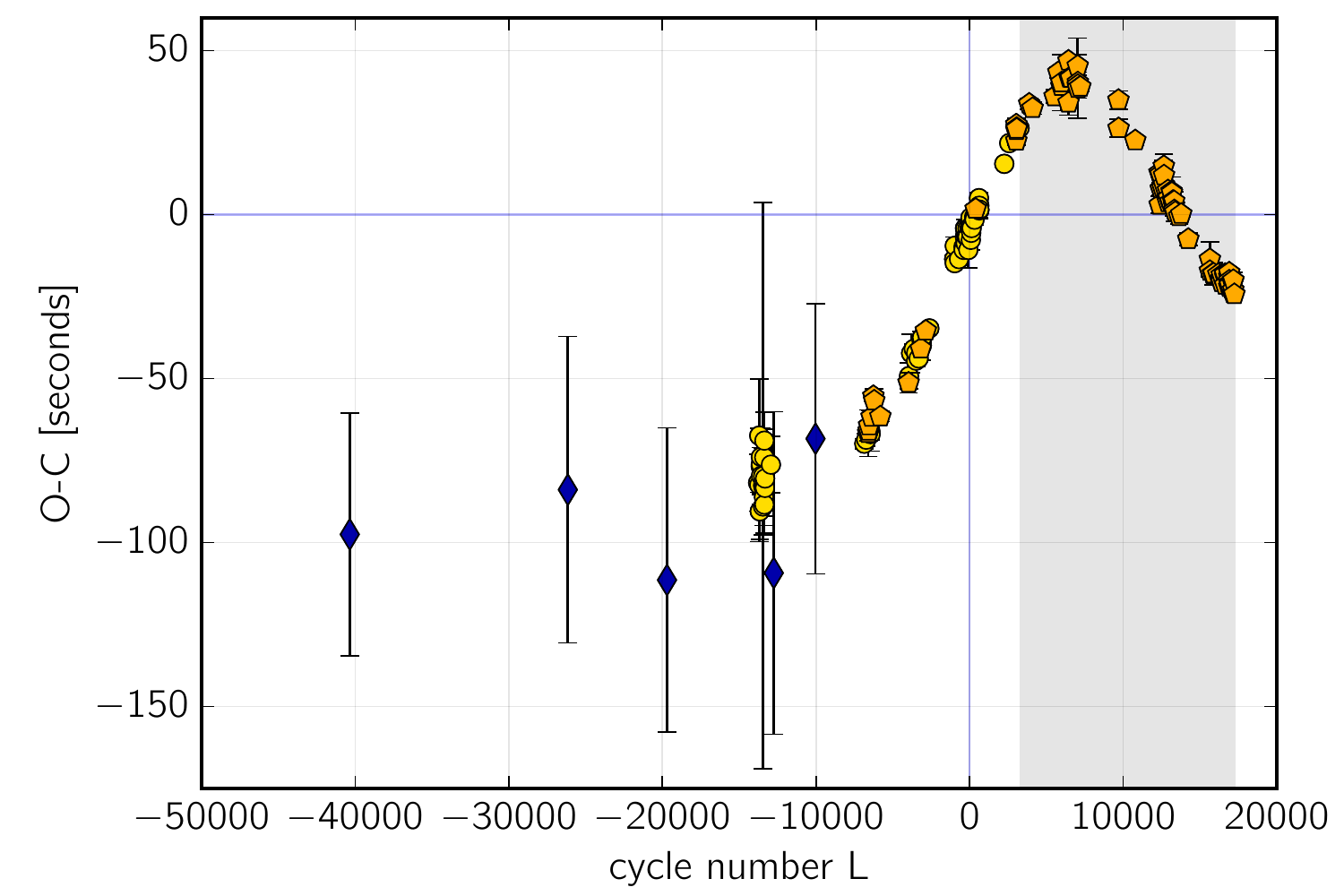}
    \caption{The (O--C) diagram w.r.t. the linear ephemeris for Dataset~B. Dark blue diamonds are for the NSVS and ASAS measurements, filled circles are for data in the up-to date literature excluding SuperWASP measurements, dark diamonds are for NSVS and ASAS data, and darker pentagons are for the new measurements in this paper (Tab.~\ref{tab:eclipses}). See the text for details.}
    \label{fig:nsvs_model}
\end{figure}

\begin{figure}
\includegraphics[angle=0, width=0.47\textwidth]{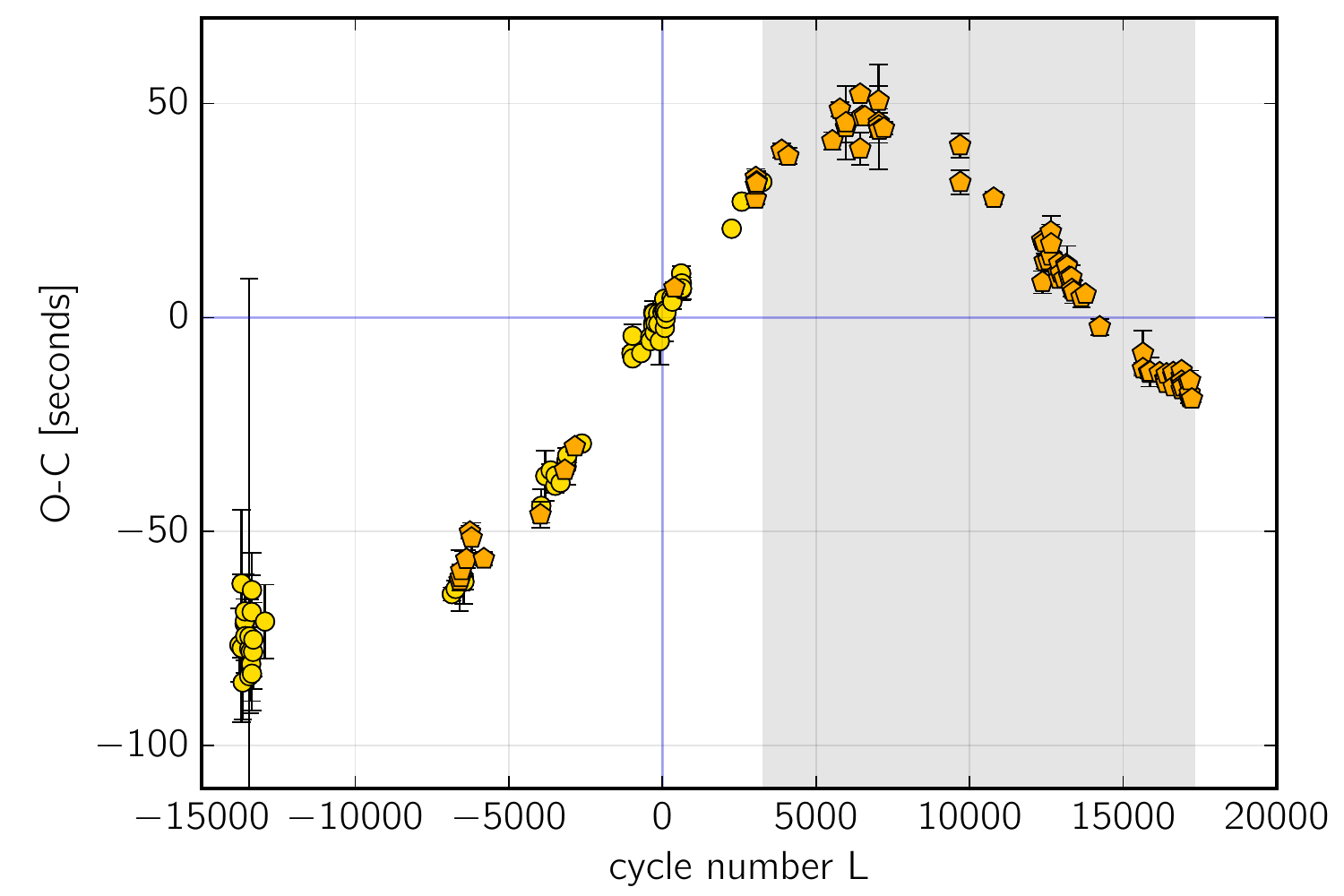}
    \caption{The (O--C) diagram w.r.t. the linear ephemeris for Dataset~C. Filled circles are for data in the up-to date literature excluding SuperWASP, NSVS and ASAS measurements, and pentagons are for the new timing data in this paper (Tab.~\ref{tab:eclipses}), see the text for details.}
    \label{fig:nsvs_best}
\end{figure}

Due to the non-homogeneous timing data which are gathered across literature and in this manuscript, we consider three datasets. Dataset~A includes all observations available to date, which encompasses
CCD observations and 5~measurements from the NSVS and ASAS archives in \cite{Beuermann2012}, SuperWASP-derived timing data \citep{Lohr2014}, as well as our new measurements listed in Table~\ref{tab:eclipses}. Due to a large scatter and uncertainties, the SuperWASP measurements are finally excluded in Dataset~B. In Dataset~C we also excluded the NSVS and ASAS measurements due to uncertain derivation of these measurements, which is discussed in Sect.~\ref{sec:noasas}. Then we subsequently analysed Datasets A, B and C individually. These particular datasets are illustrated in the (O--C) diagrams in Figs.~\ref{fig:nsvs_WASP}, \ref{fig:nsvs_model} and \ref{fig:nsvs_best}, respectively.

In all figures included in this section, we marked with  grey rectangles the time interval after the last epoch in \cite[][]{Hinse2014}. i.e. August 2012, which indicates our new measurements.  
We may expect that the orbital period of a putative third object may be constrained and that changes conclusions of \cite{Hinse2014}, who were able to see only an increase of the (O--C).

Parameters of the linear ephemeris for Dataset~A are
\[
 T_{\rm eph}(L) = \mbox{BJD}\,2455793.84004(2) + L \, 0.110374083(2),
\]
for Dataset~B the linear ephemeris is described through
\[
T_{\rm eph}(L) = \mbox{BJD}\,2455793.84005(3) + L \, 0.110374082(3),
\]
while that for  Dataset~C  is
\[
T_{\rm eph}(L) = \mbox{BJD}\,2455793.84005(3) + L \, 0.110374082(3),
\]
where $T_{\rm eph}$ stands here for the linear ephemeris of the BJD moment of the mid-eclipse of
the cycle $L$. We chose the initial epoch $T_0$ of the cycle $L=0$ roughly in the middle of the observational window, i.e., $T_0={\rm BJD}\,2455793.840061$. It is clear that the linear ephemeris is essentially the same, within the errors at the last significant digit marked in brackets. However, the SuperWASP, NSVS and ASAS points strongly deviate from apparently a quasi-sinusoidal pattern formed by (O--C) derived from more accurate measurements.

Before sampling the posterior, which is determined through the likelihood function (Eq.~\ref{eq:Lfun}), we first found the best-fitting parameters through maximising ${\cal L}$ with the genetic algorithm \citep{Charbonneau1995}. Next we ran the MCMC sampler for 512 initial conditions inside a small ball centered on this solution. We tested  chain lengths from 32,000 up to 768,000 samples. The latter very large number of samples may be considered redundant, given the acceptance fraction of $\simeq 0.35$  indicating an optimal output from the MCMC sampler \citep{Foreman2013}.


\begin{table}[t]
    \caption{Parameters of the third-body with linear ephemeris for Dataset~C (Figs.~\ref{fig:nsvs_model_resB} and \ref{fig:nsvs_mcmc}). Parameter $\Delta t_0$ is for the shift relative to the observational middle-window epoch $T_0~=~{\rm BJD}\,2455793.840061$. Total mass of the binary is 0.528\,$M_{\odot}$ \citep{Almeida2012}, see the text for details.}
    \label{tab:results}
    \centering
    \begin{tabular}{lllccc}
        \hline
        Parameter & Value & $+\sigma$ & $-\sigma$ \\
        \hline

             $K$ 
                 [s] 
                & 48.9 
                & 1.6 
                & 1.2\\
            
            $P$ 
                 [day] 
                & 3632.8 
                & 169.6 
                & 131.7\\
            
                 $x$ 
                 
                & 0 
                & 0.045 
                & 0.042\\
            
                 $y$ 
                 
                & 0.175 
                & 0.032 
                & 0.031\\
            
            $\tau$ 
                 [day] 
                & 7938.5 
                & 246.5 
                & 161.8\\

            $P_{\rm bin}$ 
                 [day] 
                & 0.110374099 
                & \ensuremath{2\times 10^{-9}} 
                & \ensuremath{3\times 10^{-9}}\\
            
            $\Delta t_0$ 
                 [day] 
                & \ensuremath{-5\times 10^{-5}} 
                & \ensuremath{2\times 10^{-5}} 
                & \ensuremath{2\times 10^{-5}}\\

            $\sigma_f$ 
                 [s] 
                & 1.8 
                & 0.2 
                & 0.2\\
            
            mass 
                [M$_\mathrm{Jup}$] 
                & 14.75 
                & 0.13 
                & 0.13\\
            
            $a$ 
                [au] 
                & 3.74 
                & 0.12 
                & 0.09\\
            
            $e$ 
                 
                & 0.175 
                & 0.012 
                & 0.003\\
            
            $\omega$ 
                [deg] 
                & 90.11 
                & 15.37 
                & 12.89\\
        \hline
    \end{tabular}
\end{table}

\begin{figure}
    \includegraphics[angle=0, width=0.47\textwidth]{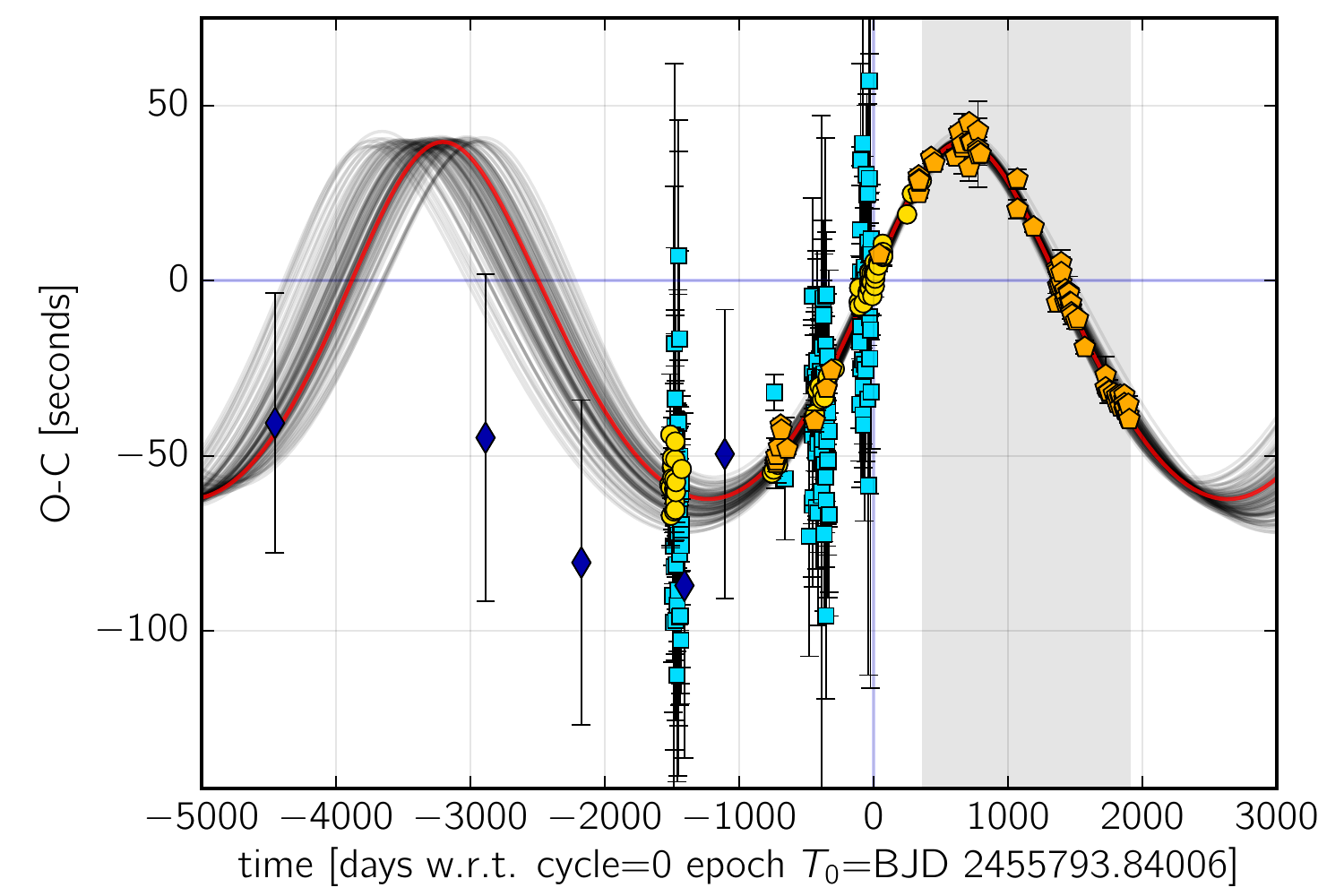}
    \includegraphics[angle=0, width=0.47\textwidth]{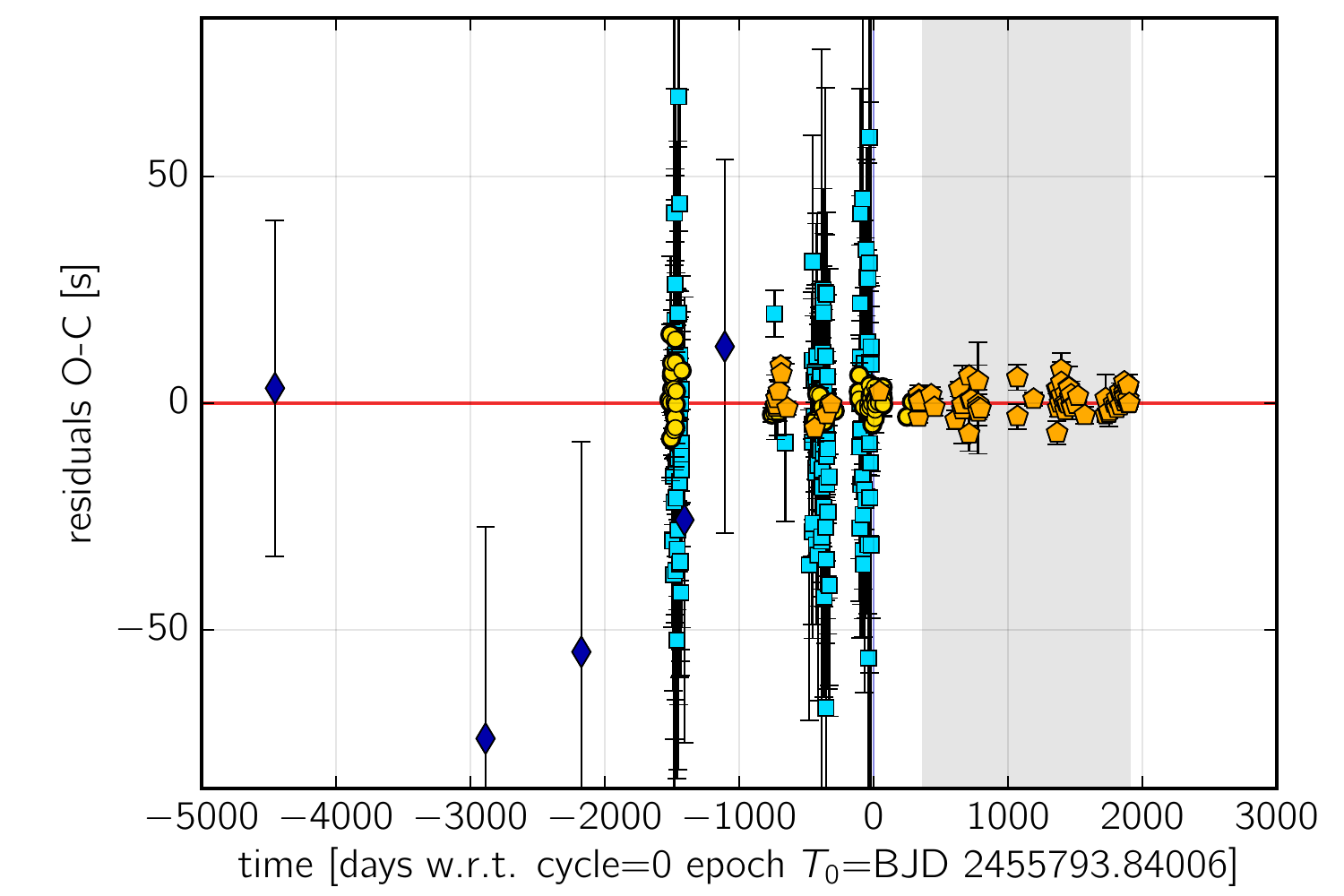}
    \caption{{\em Top panel}: the synthetic curve of the best-fitting model (red curve) to all timing data (Dataset A). Grey curves illustrate 100 randomly selected parameter samples from the MCMC posterior. {\em Bottom panel}: residuals to the best-fitting solution.}
    \label{fig:nsvs_model_WASP}
\end{figure}

\begin{figure}
    \includegraphics[angle=0, width=0.47\textwidth]{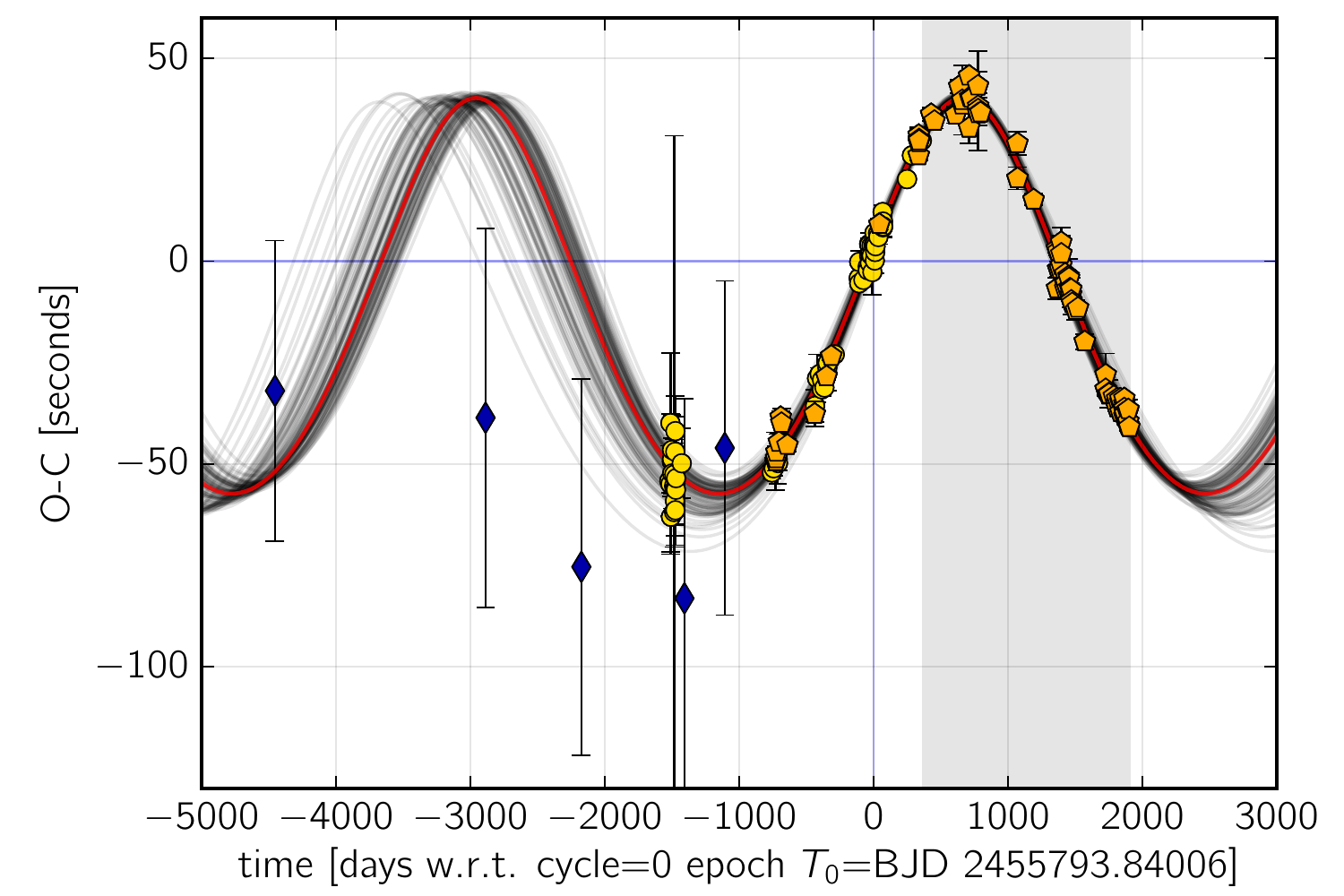}
    \includegraphics[angle=0, width=0.47\textwidth]{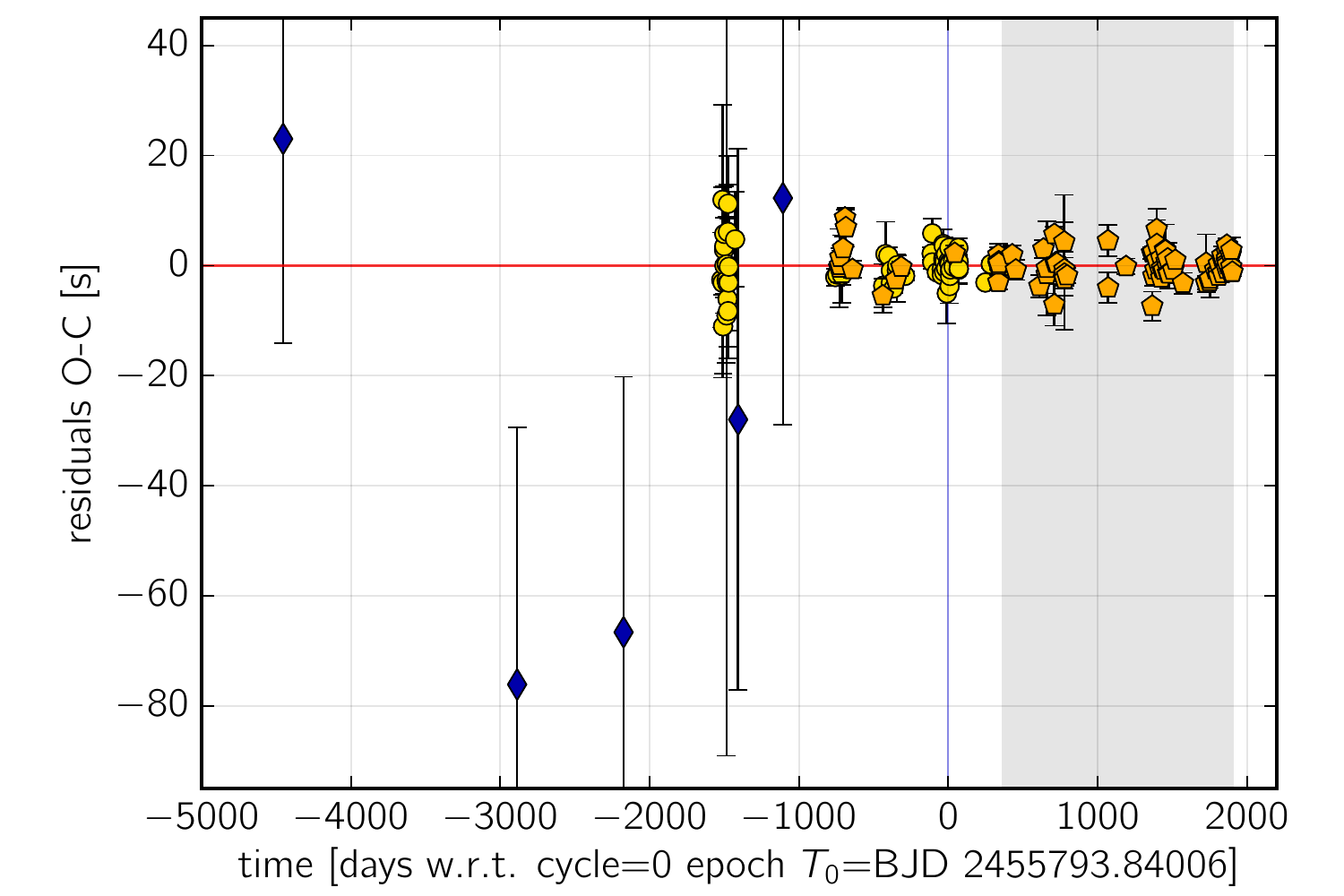}
    \caption{{\em Top panel}: the synthetic curve of the best-fitting model (red curve) to all data excluding SuperWASP data (Dataset B). Grey curves illustrate 100 randomly selected parameter samples from the MCMC posterior. {\em Bottom panel}: residuals to the best-fitting solution.}
    \label{fig:nsvs_model_res}
\end{figure}

\begin{figure}
    \includegraphics[angle=0, width=0.47\textwidth]{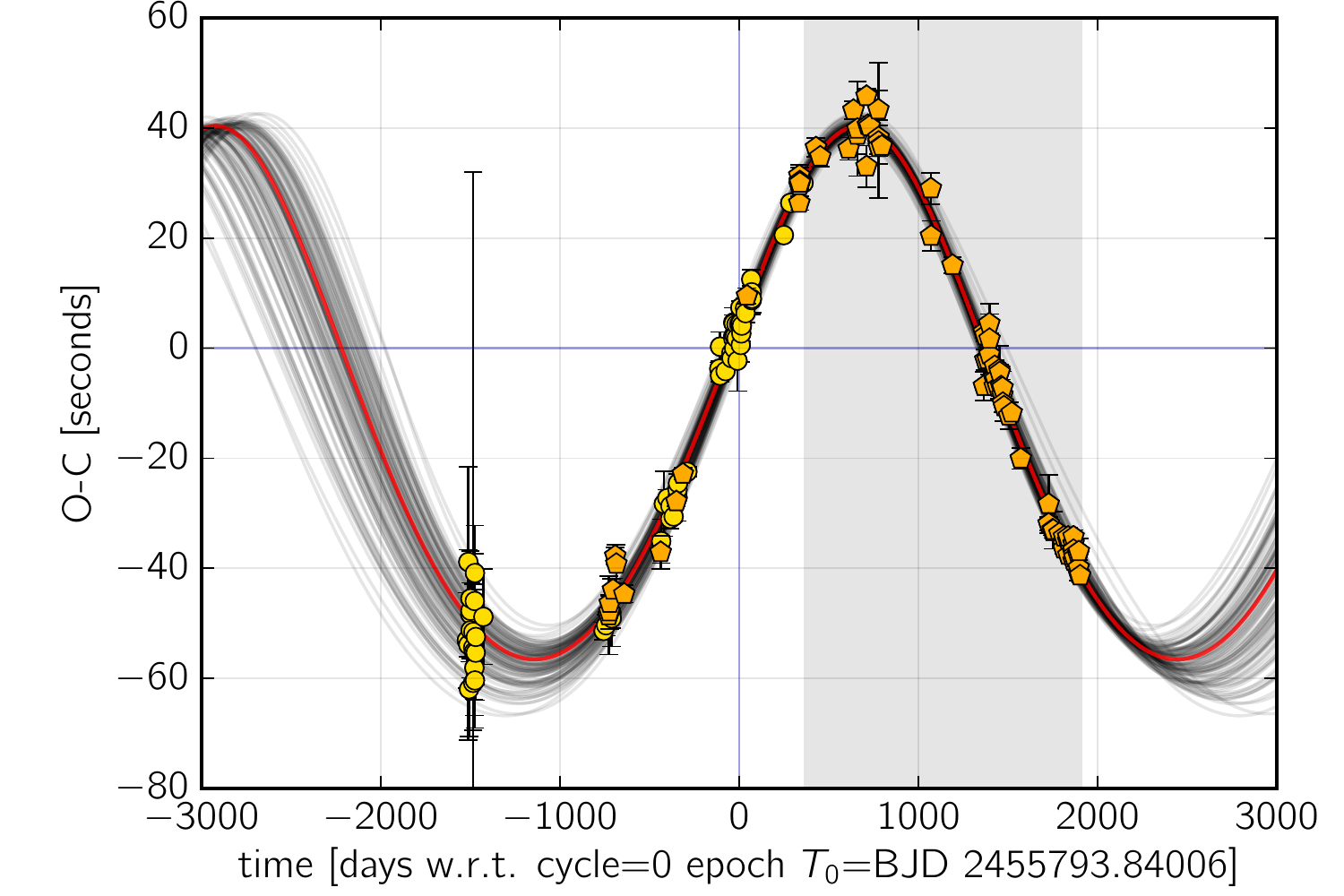}
    \includegraphics[angle=0, width=0.47\textwidth]{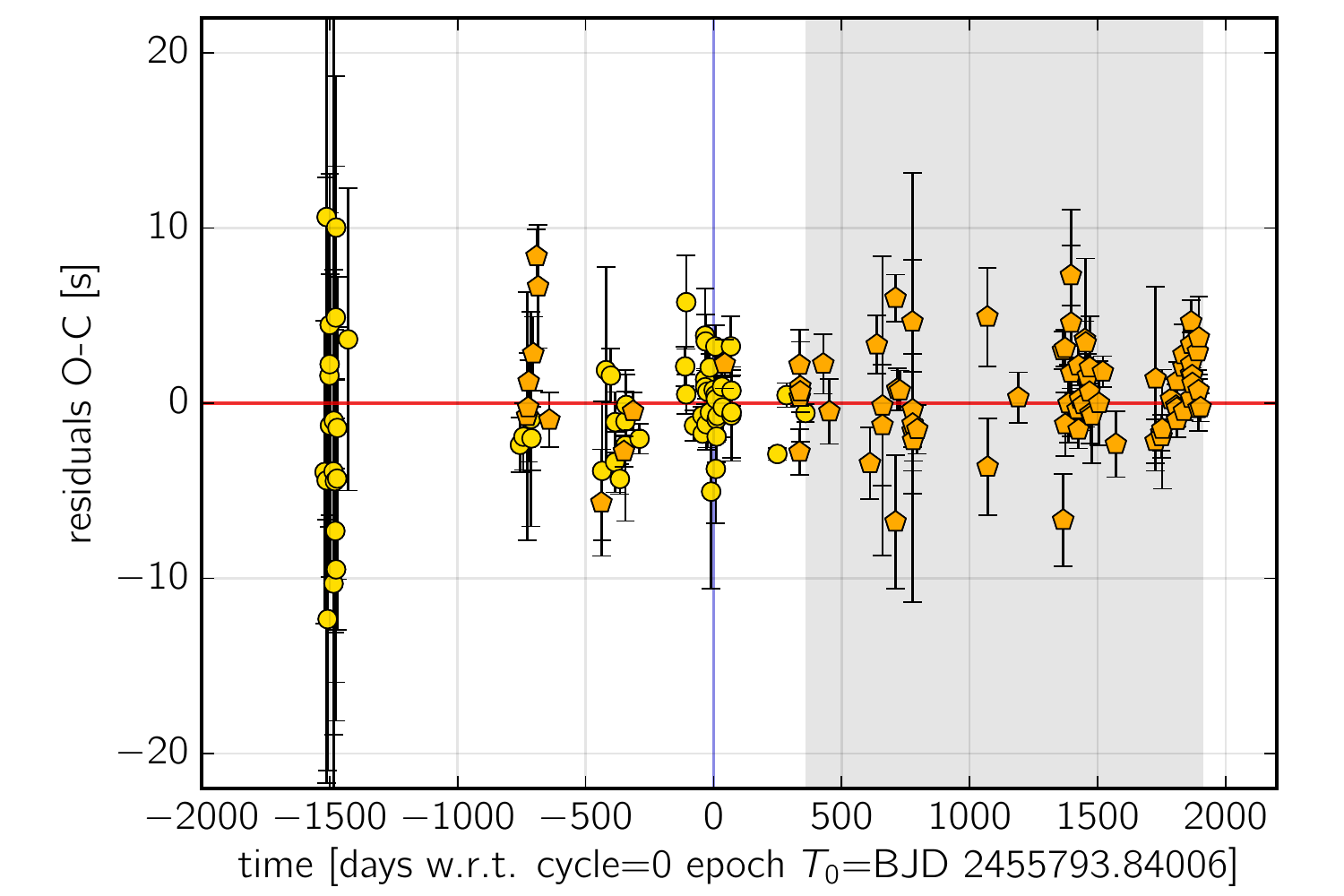}
    \caption{{\em Top panel}: the synthetic curve of the best-fitting model (red curve) to data without SuperWASP and NSVS/ASAS points (Dataset C). Grey curves illustrate 100 randomly selected parameter samples from the MCMC posterior. {\em Bottom panel}: residuals to the best-fitting solution displayed in Tab.~\ref{tab:results}.}
    \label{fig:nsvs_model_resB}
\end{figure}

\begin{figure*}
    \includegraphics[width=0.92\textwidth]{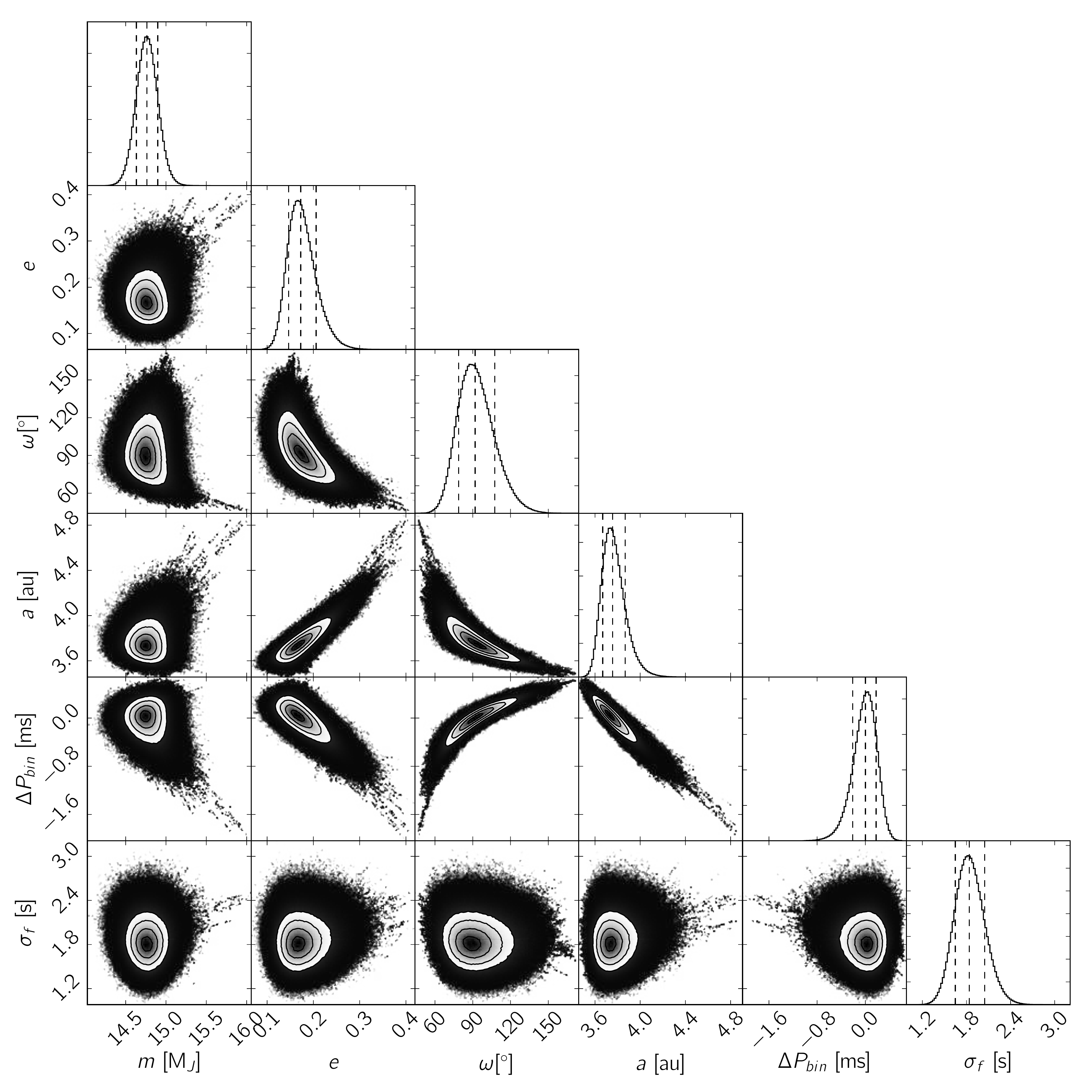} 
    \caption{One-- and two--dimensional projections of the posterior probability distributions of a few parameters inferred from the (O--C) model for Dataset~C (Tab.~\ref{tab:results} and Fig.~\ref{fig:nsvs_model_resB}). There are illustrated 32,000 samples for 512 ``walkers'' initiated in a small ball around the best-fitting model in Tab.~\ref{tab:results}. We removed about of 10\% initial, burn-out samples. Subsequent parameters are the companion mass $m$, eccentricity $e$, periastron argument $\omega$, semi-major axis $a$, a deviation of the binary period $\Delta P_{\rm bin}$ from its adopted best fitting value (in milliseconds, see Tab.~\ref{tab:results}), and the measurements uncertainty correction factor $\sigma_f$.  Contours are for the 16th, 50th and 84th percentile of samples in the posterior distribution.   See the text for details regarding parametrisation of the (O--C) model and imposed priors.}
    \label{fig:nsvs_mcmc}
\end{figure*}

 The best-fitting models and their residuals are illustrated in the top and bottom panels of Figs.~\ref{fig:nsvs_model_WASP}, \ref{fig:nsvs_model_res} and \ref{fig:nsvs_model_resB}, for A, B and C datasets, respectively. The posterior distribution is illustrated only for Dataset~C (Fig.~\ref{fig:nsvs_mcmc}), since the posteriors for Datasets A and~B are very similar, and therefore, to save space, we do not quote them. We note that the time of pericenter argument $\tau$, the binary period $P_{\rm bin}$, and the time-shift $\Delta t_0$  from the cycle $L=0$ epoch  are represented relative to the best-fitting parameters in Tab.~\ref{tab:results}, respectively, where $T_0=\mbox{BJD}\,2455793.840061$. These parameters are very close to the initial values derived with the common maximization of the likelihood function ${\cal L}$.

The posterior projections reveal relatively significant  correlations of particular model parameters, like $(K,P)$ and $(P,\tau$). However, the posterior is uni-modal with a quite strong peak. This is illustrated in Fig.~\ref{fig:nsvs_mcmc} for a few selected parameters of the (O-C) model. The MCMC sampling reveals the error floor of $\simeq 2$ seconds for models optimal in the sense defined above. Without this correction, the ``raw''  reduced $\chi_{\nu}^2 \sim 2$ indicates underestimated uncertainties. The optimal solution is represented with a red curve, and is overplotted on 100~randomly selected model curves from the MCMC sample. We found that the eccentricity of the best-fitting orbit $e \simeq 0.175$, indicating a significantly skewed (O--C) curve, and a relatively large semi-amplitude of the LTT signal $K\simeq 50$~s, rule out pericenter precession of the orbit, following \cite{Beuermann2012}.

Due to apparently random residuals with a rms $\simeq 10$~seconds, which is almost equal to the mean of the rescaled uncertainties, we did not analyse models with additional parameters, such as the parabolic ephemeris \citep{Hinse2014} or even a putative second companion \citep{Almeida2013}. The most simple, 1-companion model with the linear ephemeris, does not exhibit systematic, long-term changes of the (O--C) superimposed on the quasi-sinusoidal variation. Secular changes of the orbital period  should not be expected for such a detached binary \citep{Beuermann2012}.

To infer the companion mass from the third-body model parameters listed in Tab.~\ref{tab:results}, we used the stellar masses as $M_1 = 0.419~M_\odot$ and $M_2 = 0.109~M_\odot$ for the primary and the secondary, respectively, following \cite{Almeida2012}. The best-fitting orbital period of $P \simeq 3600$~days implies the minimal mass of $\sim 15$ Jupiter masses (when the orbits are co-planar), that is in the brown dwarf.  \corr{For the ratio of orbital periods  $P/P_{\rm bin}\sim 4\times 10^4$, the triple system is highly hierarchical. Obviously, the brown dwarf has a stable orbit, which is two orders of magnitude wider than the stability limit $\simeq 0.2\,a_{\rm bin}$ (roughly $\sim 0.01$~au for the \nsvs{} binary) expected for circumbinary companions, if the binary eccentricity $e_{\rm bin} \simeq 0$ \citep[e.g.,][their Table 7]{Holman1999}.}
\corr{In such a case, the brown dwarf eccentricity $\sim 0.2$ has a negligible impact on the stability.}

The third-body parameters determined here substantially differ from previous estimates.  For instance, \cite{Beuermann2012} reported the orbital period unconstrained between 20 and 70~years with eccentricity $e \simeq 0.50$ for a $20$~yrs orbit, since their data did not cover the (O--C) maximum revealed here. We determine the semi-amplitude of the LTT signal being roughly twice larger than that stated in \cite{Hinse2014}. The amplitude of 
(O--C) is one of crucial parameters needed to estimate the energy required to support the Applegate cycles \citep[e.g.,][]{Volschow2016}.

\section{Discussion}
\label{sec:discussion}

\subsection{The NSVS and ASAS timing data}
\label{sec:noasas}
We would like to comment on the five earliest points from the literature, i.e. one timing measurement from the NSVS \citep{Wozniak2004nsvs} survey and four measurements from the ASAS survey \citep{Pojmanski1997}, all five presented by \citet{Beuermann2012}. 

It may be observed that these data exhibit large uncertainties and they strongly deviate from the models shown in this work. The exposure time is 80~seconds for NSVS survey, and as long as 180 seconds for the ASAS light curves. Moreover, to derive eclipses, one must collect measurements by folding photometric points spanning over a whole year. 

Therefore we decided to re-analyse the source photometric data to derive the timings in an independent way. We downloaded the source light curves from the publicly available  NSVS\footnote{http://skydot.lanl.gov/} and ASAS\footnote{http://www.astrouw.edu.pl/asas/}  archives. The NSVS light curve spans epochs between 1999 and 2000, while the ASAS observations span over a few years between 2003 and 2008. The photometric NSVS measurements
timestamps are given in MJD UTC, while the ASAS measurements in HJD UTC. Therefore we recomputed the observation moments to the standard Solar system barycenter BJD time-scale with a procedure developed by \cite{Eastman2010}. The data have been divided into $~\sim 1$~year intervals and phase-folded with the orbital period of the binary derived from the most recent linear ephemeris. We know that this period is determined with an uncertainty to a few parts of millisecond, and assuming that the linear ephemeris is valid, we may use this new, fixed estimate.

In the next step we attempted to fit the function representing primary eclipses, as proposed in \cite{Beuermann2012}. 
We performed two experiments. In the first one, we fitted the whole set of parameters. In the second experiment, we constructed the mean, synthetic light-curve from the best 71~eclipses derived with the standard method. Then we fitted only some of the model eclipse parameters as free. 

Unfortunately, all these experiments resulted in variable mid-eclipses estimates spread over 90~seconds. Moreover, for the ASAS data, the O-C of our timing moments are systematically $\sim 60$~second {\em earlier} than  the measurements listed in \cite{Beuermann2012}.

The obtained formal errors from the model curve fits were on the level of $\pm$35-42 seconds and are similar to the (O-C) deviations from the linear ephemeris. Also the folding of photometric data for a year (which is around 3300 binary cycles) introduces a systematic shift of the mid-eclipses, in accord with the local (O--C) trend. It could  be as large as $\sim 20-50$ seconds. 

We conclude that the NSVS and ASAS mid-eclipse measurements  are not very useful for the O-C analysis od such a short period binary. Fortunately, the parameters of our O-C models derived for each of the three data-sets are similar. We believe that  the final fit shown in Tab.~\ref{tab:results}, derived for Dataset~C, does indeed represent a reliable solution. 

\subsection{The Applegate mechanism of the (O--C)}
\label{sec:applegate}
%
%
For compact binaries, like \nsvs{}, magnetic activity of the less massive component may trigger Solar-like cycles and reshape the internal structure of this star \citep[e.g.,][]{Applegate1992,Lanza1998,Brinkworth2006}. This leads to changes of the mutual gravitational field and oscillations of the orbital period. A common problem for this origin of the (O--C) variations is insufficient energy budget of the secondary required to change its quadrupole moment $Q$. Therefore, the Applegate cycles are usually dismissed in the literature as a possible  explanation of cyclic variations of the (O--C) observed in a number of \corr{PCEBs}. There are, however, more detailed and improved models of the Applegate mechanism, which modify the energy requirements \citep{Lanza1998,Lanza2006,Brinkworth2006}. Recently, \cite{Volschow2016} considered a few variants of the \citet{Brinkworth2006} formulation that generally takes into account more realistic stellar density profiles. They analysed a sample of 15 compact \corr{PCEBs}, including \nsvs{}, and found that only for four systems in the sample, the magnetic cycles may be responsible for the (O--C) behaviour. For \nsvs{} the relative energy $\Delta E/E_{\rm sec}$ required to trigger the measured (O--C) should  be between $\sim 5.4$ for the ``classic'' Applegate model and $\simeq 100$ for an advanced model of the stellar density profile (for a constant density profile, the ratio is 2 orders of magnitude larger, $\sim 3000$). 

 The up-to date (O--C) analysed in this paper implies a substantial change of the semi-amplitude $K$ and the variation period, we recalculated estimates  of the energy budget $\Delta E/E_{\rm sec}$ for \nsvs{} given in \cite{Volschow2016}. We recomputed this value in accord with their Eq. 7, following \cite{Tian2009},  for canonical models in \citet{Applegate1992} as well for the modified Applegate mechanism in \citet[][and references therein]{Lanza1998,Lanza2006}. We used also data for the secondary component from their Tab.~1. 

Adopting the secondary radius $R_2=0.162~R_{\odot}$, mass $M_2 = 0.109~M_{\odot}$, orbital separation $a=0.80~R_{\odot}$, and the effective temperature $T =2550$~K we found that $\Delta E/E_{\rm sec}$ $\simeq 11$. We computed the period change relative to the binary period
\[
 \frac{\Delta P}{P_{\rm bin}} = 4\pi  \frac{K}{P} \simeq 2\times 10^{-6}, 
\]
with the semi-amplitude $K \simeq 49$~s and (O-C) oscillation $P \simeq 9.95$~yrs (modulation period) as displayed in Tab.~\ref{tab:results}. The quadrupole period variation $\Delta Q$ needed to drive the modulation of the orbital period \citep{Lanza1999} is:
\[
 \frac{\Delta P}{P_{\idm{bin}}} = -9 \frac{\Delta Q}{M_{\rm bin} a_{\rm bin}^2}, 
\]
where $M_{\rm bin}$ and $a_{\rm bin}$ are the binary mass and the semi-major axis, respectively. For \nsvs{}, we obtain the magnitude of $\Delta Q \simeq 10^{47} {\rm g\,cm}^2$. 

The updated $\Delta E/E_{\rm sec}$ is more than two times larger than the value in \citet{Volschow2016} for the genuine Applegate model which, in accord with their analysis, tends to underestimate the energy ratio. For other variants, based on the \citet{Brinkworth2006} formulation, and realistic stellar  density profiles,  the Applegate modulations are even less probable, since the prescribed energy budget is then by 1-2 orders of magnitude too small, as shown in \citet[their Tab.~4]{Volschow2016}. 

In accord with the alternative generalisation of the Applegate mechanism by \citet{Lanza1998,Lanza2006}, taking into account additional factor of the Lorenz force, the magnetic cycles may operate with a~fraction of the energy required by the original Applegate model. Yet the lower limit of the calculated  ${\Delta E}/{E_{\rm sec}} \sim 11$ factor even for this scenario seems to be so large that one can safely conclude that the Applegate mechanism and its generalizations proposed by \citet{Brinkworth2006}, \citet{Lanza2006} and \citet{Volschow2016} are not a credible explanations of the (O--C) variability in the \nsvs{} binary.

%
%

\section{Conclusions}
\label{sec:conclusions}

Our new set of light-curves of the \nsvs{} binary substantially extends the archived list of eclipse timing. For the first time our new data cover the maximum of the (O--C) w.r.t. the linear ephemeris, 
covering almost one full cycle of a  quasi-sinusoidal modulation, and making it possible to put constrains on previous \corr{LTT} models aimed at explaining the (O--C) behaviour of this system  \citep{Beuermann2012, Almeida2013, Hinse2014}.

In accord with the third-body hypothesis, the observed (O--C) variations in the \nsvs may be explained by the presence of a single companion with a minimal mass in the brown dwarf mass range (14.7~Jupiter masses), in a moderately eccentric orbit with eccentricity $\simeq 0.2$, and the orbital period of $\sim$ 10 years. We found that parameters of this third-body within our best model are relatively well constrained through the present data. The residuals do not indicate \corr{any significant} secular trends which could appear due to dissipative phenomena in the binary (like mass transfer, magnetic braking and gravitational radiation). We note that \cite{Beuermann2012} and \cite{Hinse2014}  reported such trends due to much shorter observational window that did not cover the  maximum of the (O--C) shown in this work.  

Therefore,  a 1-companion model  may be the most reliable explanation of the  \nsvs{} (O--C). Taking into account the updated amplitude of LTT of $K \simeq 50$~s and its period of $\simeq 10$~yr, the alternative hypothesis -- the Applegate mechanism -- does not seem to be sufficiently effective to produce such changes.  In accord with a very recent analysis by \cite{Volschow2016}, the energy required to trigger the Applegate cycles in the secondary companion should be 10--100 larger than its nuclear energy. Moreover, the relatively large $K$ and the the shape of (O--C) yielding the third-body orbit eccentricity of $\simeq 0.17$, rule out also the orbital precession as a plausible (O--C) variations mechanism. 

\corr{The presence of a massive companion in moderately eccentric orbit around the evolved, compact binary would be not necessarily unusual on the grounds of the planet formation theory. Many scenarios are possible, regarding both first generation planets (companions) that survived the Common-Envelope (CE) phase, as well as emerged in a protoplanetary disc formed from  the stellar matter ejected during the CE phase, as second generation planets \citep[e.g.,][]{Veras2011,Veras2012,PortegiesZwart2013,Bear2014,Volschow2014,Kostov2016b,Veras2017}. In the first case, the best-fitting orbital elements of \nsvs{} may be used as the border conditions required to reconstruct the binary evolution, as shown by \cite{PortegiesZwart2013}.}

We do not analyse other possibilities of the (O--C) variability, like the mass transfer, orbital precession, magnetic braking or gravitational radiation, which are usually refuted for this class of binaries.

We should stress, however, that \corr{the third-body hypothesis investigated for} a number of close and evolved PCEBs reported in the literature, remains uncertain in most cases. A very discouraging example of this kind is the Cataclysmic Variable (CV) polar, HU Aqr \citep{Schwope1993,Schwarz2009,Qian2011,Hinse2012,Gozdziewski2012,Schwope2014,Bours2014,Gozdziewski2015}. The apparently quasi-sinusiodal (O--C) variations with a full amplitude of $\sim 60$~s observed for almost 20~years till 2012, has changed to a strong, secular trend that deviates by 180~s from the linear ephemeris to date. Two and three planet models of this system are strongly unstable,
unless we consider an exotic system of three $5-6~M_{\rm Jup}$ mass planets, with the middle one revolving in a retrograde orbit. Such a system may be stable for at least 1~Gyr \citep{Gozdziewski2015}. 
\corr{Similarly, the (O-C) of HW~Vir  interpreted through 5:2~MMR configuration of two Jovian planets \citep{Beuermann2012b} are not constrained, regarding the outermost planet and its mass. The (O-C) for other PCEBs, like NY~Vir \citep{Qian2012b,Lee2014}, QS Vir \citep{Horner2013}, UZ~For \citep{Potter2011} could be formally explained with the resonant 2-planet systems, yet none of them has been found stable.} An exception is
the NN Ser with low-mass, Jovian planets close to 2:1~mean motion resonance, which is well documented nad has passed  \corr{so far all tests of the planetary nature of the  (O--C) \citep{Beuermann2010,Marsh2014,Volschow2014}.}
\corr{
A few other PCEBs, like V471 Tau \citep{Hardy2015}, V470~Cam \citep{Qian2013}, RR~Cae \citep{Qian2012a} might host single-companions, see \cite{Almeida2013} and \cite{Zorotovic2013} listing such binaries with their astrophysical characteristics.}

\corr{
Observations of these systems is still timely since the long-term, hardly predictable (O-C) are typically known for a fraction of the longest putative orbital periods. There are open problems regarding the PCBEs, like the formation of putative companions as first- or second-generation planets, orbital architectures and stability of hypothetical multiple-companion configurations, the presence of mechanisms alternative or coexisting with the Applegate and Lanza-Rodon\'o cycles and the LTT effect. Therefore, while the planetary hypothesis of the (O--C) observed for \nsvs{} cannot be yet definite, our observations and new data may contribute more light on the unresolved astrophysical questions. For instance, the (O-C) amplitude constrains the energy required to trigger magnetic cycles of the M-dwarf component.}

Additional, long-term timing observations of the \nsvs binary are required. Being relatively bright, the \nsvs{} system  may be systematically monitored, as we show here, with  $\sim 1$~m class telescopes. During next 2-3 years, the third-body model and the eclipse ephemeris can be verified due to (O--C) approaching the nearby minimum (see Fig.~\ref{fig:nsvs_best} for our prediction). 

\section*{Acknowledgements}

We thank to the anonymous reviewer for his/her critical and informative comments which improved and this work. This work has been supported by The Scientific and Technological Research Council of Turkey (TUBITAK), through project number 114F460 (IN, AS, HE). This work has been supported by Polish National Science Centre grant DEC-2011/03/D/ST9/00656 (AS, KK, M\.Z). We thank the team of TUBITAK National Observatory (TUG) for a partial support in using T100 telescope with project number TUG T100-63.  We also wish to thank Adiyaman University Observatory (Turkey) and the Skinakas Observatory (Heraklion, Greece) teams. K.G. thanks the staff of the Poznan Supercomputer and Network Centre (PCSS) for the support and computational resources (grant 195).

\facility{Skinakas:1.3m}
\software{Python, IRAF, SExtractor, emcee \citep{Foreman2013}}

\section*{Appendix: Additional figures and tables}
\label{sec:appendix}

\begin{table}
\caption{\nsvs observations log: starting date of observations, cycle,
eclipse type (1 for primary, 2 for secondary), filter, exposure time,
readout time and observatory code: KRK -- the Astronomical Observatory of
the Jagiellonian University, SUH -- the Mt.  Suhora Observatory, SKO --  the
Skinakas Observatory, TUG -- the TUBITAK National Observatory, ADYU60 -- the
Adiyaman University Observatory. Exp $\equiv$ exposure time in seconds,
Rd $\equiv$ redout time in seconds, Obs. $\equiv$ observatory. 
}
\label{tab:observations}
\begin{tabular}{ccccccc}
\hline
Date & $L$ & Type & Filter & Exp & Rd & Obs. \\ 
\hline
  2009-08-21 & 7167.5 & 2 & BG40 & 10 & 2.2 & KRK \\ 
  2009-08-25 & 7204.0 & 1 & BG40 & 12 & 2.3 & KRK \\ 
  2009-08-27 & 7223.0 & 1 & BG40 & 10 & 2.2 & KRK \\ 
  2009-09-14 & 7386.0 & 1 & W-light & 12 & 3.0 & SUH \\ 
  2009-09-27 & 7503.0 & 1 & BG40 & 12 & 2.2 & KRK \\ 
  2009-10-03 & 7557.0 & 1 & BG40 & 12 & 2.3 & KRK \\ 
  2009-11-16 & 7955.0 & 1 & W-light & 10 & 3.0 & SUH \\ 
  2010-06-08 & 9797.0 & 1 & BG40 & 12 & 5.0 & KRK \\ 
  2010-09-03 & 10593.0 & 1 & R &  8 & 2.8 & SUH \\ 
  2010-10-09 & 10918.0 & 1 & R & 12 & 2.8 & SUH \\ 
  2011-10-02 & 14162.0 & 1 & R &  5 & 2.4 & SUH \\ 
  2012-07-20 & 16808.0 & 1 & R &  5 & 12.0 & SKO \\ 
  2012-07-20 & 16808.5 & 2 & R &  5 & 12.0 & SKO \\ 
  2012-07-21 & 16817.0 & 1 & R &  5 & 12.0 & SKO \\ 
  2012-07-23 & 16835.0 & 1 & BG40 & 10 & 2.4 & SUH \\ 
  2012-07-23 & 16835.5 & 2 & BG40 & 10 & 2.4 & SUH \\ 
  2012-10-21 & 17650.0 & 1 & BG40 & 12 & 2.0 & KRK \\ 
  2012-11-14 & 17867.0 & 1 & BG40 & 10 & 2.0 & KRK \\ 
  2013-04-22 & 19301.0 & 1 & R & 12 & 4.8 & SUH \\ 
  2013-05-18 & 19545.0 & 1 & R & 10 & 2.4 & SUH \\ 
  2013-06-09 & 19744.0 & 1 & W-light & 15 & 4.5 & SUH \\ 
  2013-06-09 & 19744.5 & 2 & W-light & 15 & 4.5 & SUH \\ 
  2013-07-30 & 20206.0 & 1 & W-light & 10 & 5.0 & SUH \\ 
  2013-07-30 & 20206.5 & 2 & W-light & 10 & 5.0 & SUH \\ 
  2013-08-06 & 20269.0 & 1 & R & 10 & 4.6 & SUH \\ 
  2013-08-16 & 20360.0 & 1 & R & 15 & 4.8 & SUH \\ 
  2013-10-04 & 20803.0 & 1 & R &  3 & 4.8 & SUH \\ 
  2013-10-05 & 20812.0 & 1 & R & 10 & 3.1 & SUH \\ 
  2013-10-05 & 20812.5 & 2 & R & 10 & 3.1 & SUH \\ 
  2013-10-05 & 20813.0 & 1 & R & 10 & 3.1 & SUH \\ 
  2013-10-07 & 20830.0 & 1 & R & 15 & 4.8 & SUH \\ 
  2013-10-23 & 20975.0 & 1 & BG40 & 12 & 3.0 & KRK \\ 
  2014-07-24 & 23458.0 & 1 & W-light &  5 & 13.9 & TUG \\ 
  2014-07-25 & 23467.0 & 1 & W-light &  5 & 13.9 & TUG \\ 
  2014-11-22 & 24553.0 & 1 & W-light &  5 & 13.7 & TUG \\ 
  2015-05-16 & 26132.0 & 1 & W-light & 15 & 1.0 & ADYU60 \\ 
  2015-05-16 & 26141.0 & 1 & W-light & 15 & 1.0 & ADYU60 \\ 
  2015-05-24 & 26205.0 & 1 & W-light &  5 & 14.0 & TUG \\ 
  2015-05-24 & 26213.0 & 1 & W-light &  3 & 14.0 & TUG \\ 
   \hline
\end{tabular}
\end{table}

\addtocounter{table}{-1}

\begin{table}
\centering
\caption{\nsvs Observations log: {\bf Tab.3~cont.}}
\label{tab:observations}
\begin{tabular}{ccccccc}
\hline
Date & $L$ & ET & Filter & Exp. & R & Obs. \\ 
\hline
  2015-06-06 & 26331.0 & 1 & W-light & 10 & 1.0 & ADYU60 \\ 
  2015-06-16 & 26413.0 & 1 & W-light &  5 & 13.9 & TUG \\ 
  2015-06-17 & 26422.0 & 1 & W-light &  5 & 14.0 & TUG \\ 
  2015-06-17 & 26422.5 & 2 & W-light &  7 & 14.1 & TUG \\ 
  2015-07-11 & 26648.0 & 1 & W-light & 15 & 1.0 & ADYU60 \\ 
  2015-07-15 & 26683.0 & 1 & W-light &  5 & 20.0 & TUG \\ 
  2015-07-15 & 26684.0 & 1 & W-light &  5 & 20.0 & TUG \\ 
  2015-07-21 & 26730.0 & 1 & W-light & 10 & 1.0 & ADYU60 \\
  2015-08-02 & 26846.0 & 1 & W-light & 10 & 1.0 & ADYU60 \\ 
  2015-08-09 & 26911.0 & 1 & W-light & 15 & 2.0 & ADYU60 \\ 
  2015-08-12 & 26937.5 & 2 & W-light &  5 & 12.9 & TUG \\ 
  2015-08-20 & 27009.0 & 1 & W-light & 10 & 6.0 & ADYU60 \\ 
  2015-08-27 & 27073.0 & 1 & W-light & 15 & 5.0 & ADYU60 \\ 
  2015-08-28 & 27082.0 & 1 & W-light & 10 & 1.0 & ADYU60 \\ 
  2015-08-30 & 27099.0 & 1 & W-light & 16 & 1.0 & ADYU60 \\ 
  2015-09-05 & 27154.0 & 1 & W-light & 15 & 5.0 & ADYU60 \\ 
  2015-10-03 & 27408.0 & 1 & W-light & 20 & 5.0 & ADYU60 \\ 
  2015-10-18 & 27543.0 & 1 & W-light & 10 & 1.0 & ADYU60 \\ 
  2015-12-08 & 28004.0 & 1 & W-light & 15 & 1.0 & ADYU60 \\ 
  2016-05-11 & 29411.5 & 2 & W-light &  7 & 14.0 & TUG \\ 
  2016-05-12 & 29412.0 & 1 & W-light &  7 & 14.1 & TUG \\ 
  2016-06-02 & 29611.0 & 1 & W-light & 15 & 1.0 & ADYU60 \\ 
  2016-06-03 & 29620.0 & 1 & W-light & 15 & 1.0 & ADYU60 \\ 
  2016-06-06 & 29647.0 & 1 & W-light & 15 & 1.0 & ADYU60 \\ 
  2016-07-11 & 29956.0 & 1 & W-light & 10 & 14.9 & TUG \\ 
  2016-07-30 & 30135.0 & 1 & W-light & 15 & 1.0 & ADYU60 \\ 
  2016-08-02 & 30163.0 & 1 & W-light & 15 & 1.0 & ADYU60 \\ 
  2016-08-03 & 30172.0 & 1 & W-light & 15 & 1.0 & ADYU60 \\ 
  2016-08-04 & 30180.0 & 1 & W-light & 15 & 1.0 & ADYU60 \\ 
  2016-08-28 & 30397.0 & 1 & W-light &  7 & 13.7 & TUG \\ 
  2016-08-28 & 30399.0 & 1 & W-light &  7 & 13.7 & TUG \\ 
  2016-08-29 & 30408.0 & 1 & W-light & 10 & 14.1 & TUG \\ 
  2016-09-26 & 30660.0 & 1 & W-light & 15 & 1.0 & ADYU60 \\ 
  2016-09-27 & 30669.0 & 1 & W-light & 15 & 1.0 & ADYU60 \\ 
  2016-09-27 & 30669.0 & 1 & W-light & 10 & 14.1 & TUG \\ 
  2016-09-27 & 30670.0 & 1 & W-light & 15 & 1.0 & ADYU60 \\ 
  2016-09-27 & 30670.0 & 1 & W-light & 10 & 14.1 & TUG \\ 
  2016-10-01 & 30705.0 & 1 & W-light & 15 & 1.0 & ADYU60 \\ 
  2016-10-02 & 30714.0 & 1 & W-light & 15 & 1.0 & ADYU60 \\ 
  2016-10-23 & 30904.0 & 1 & W-light & 15 & 1.0 & ADYU60 \\ 
  2016-10-26 & 30931.0 & 1 & W-light & 10 & 14.1 & TUG \\ 
  2016-10-26 & 30931.0 & 1 & W-light & 15 & 1.0 & ADYU60 \\ 
  2016-10-27 & 30941.0 & 1 & W-light & 15 & 1.0 & ADYU60 \\ 
  2016-11-03 & 31004.0 & 1 & W-light & 10 & 14.1 & TUG \\ 
   \hline
\end{tabular}
\end{table}


\begin{table}
\centering
\caption{List of the \nsvs eclipse times from the literature as well as new
 measurements. The data point number, time of the minimum 
with its error
 and the type of the eclipse (1 for primary, 2 for secondary) and references are
 given. References correspond to the following papers: (1) \citet{Wils2007},
 (2) \citet{Beuermann2012}, (3) \citet{Kilkenny2012}, (4) \citet{Almeida2013}, 
 (5) the Astronomical Observatory of the Jagiellonian University (this work), (6) the Mt. Suhora Observatory (this work), (7) the Skinakas Observatory (this work),
 (8) the TUBITAK National Obs. (this work), (9) the Adiyaman Univ. Obs. (this work).
Data from \cite{Lohr2014} (see the astro-ph version) 
are not included in this Table. T is for eclipse type.} 
\label{tab:eclipses_all}
\begin{tabular}{crccccc}
  \hline
 & Cycle & BJD & Error[d] & Err[s] & T & Ref. \\ 
  \hline
1 & $-$26586.0 & 2451339.803273 & 0.000429 & 37 & 1 & 2 \\ 
  2 & $-$12390.0 & 2452906.673899 & 0.000541 & 47 & 1 & 2 \\ 
  3 & $-$5931.0 & 2453619.579776 & 0.000537 & 46 & 1 & 2 \\ 
  4 & 0.0 & 2454274.208800 & 0.000100 & 9 & 1 & 1 \\ 
  5 & 72.0 & 2454282.155900 & 0.000200 & 17 & 1 & 1 \\ 
  6 & 73.0 & 2454282.266100 & 0.000200 & 17 & 1 & 1 \\ 
  7 & 108.0 & 2454286.129100 & 0.000100 & 9 & 1 & 1 \\ 
  8 & 172.0 & 2454293.193200 & 0.000100 & 9 & 1 & 1 \\ 
  9 & 180.0 & 2454294.076200 & 0.000100 & 9 & 1 & 1 \\ 
  10 & 181.0 & 2454294.186600 & 0.000100 & 9 & 1 & 1 \\ 
  11 & 190.0 & 2454295.179900 & 0.000100 & 9 & 1 & 1 \\ 
  12 & 316.0 & 2454309.087000 & 0.001000 & 86 & 1 & 1 \\ 
  13 & 317.0 & 2454309.197300 & 0.000100 & 9 & 1 & 1 \\ 
  14 & 325.0 & 2454310.080400 & 0.000100 & 9 & 1 & 1 \\ 
  15 & 362.0 & 2454314.164200 & 0.000100 & 9 & 1 & 1 \\ 
  16 & 380.0 & 2454316.150900 & 0.000100 & 9 & 1 & 1 \\ 
  17 & 397.0 & 2454318.027400 & 0.000100 & 9 & 1 & 1 \\ 
  18 & 406.0 & 2454319.020600 & 0.000100 & 9 & 1 & 1 \\ 
  19 & 407.0 & 2454319.131200 & 0.000100 & 9 & 1 & 1 \\ 
  20 & 443.0 & 2454323.104500 & 0.000100 & 9 & 1 & 1 \\ 
  21 & 452.0 & 2454324.097900 & 0.000100 & 9 & 1 & 1 \\ 
  22 & 832.0 & 2454366.040100 & 0.000100 & 9 & 1 & 1 \\ 
  23 & 1018.0 & 2454386.569297 & 0.000569 & 49 & 1 & 2 \\ 
  24 & 3737.0 & 2454686.676900 & 0.000477 & 41 & 1 & 2 \\ 
  25 & 6914.0 & 2455037.335341 & 0.000018 & 2 & 1 & 2 \\ 
  26 & 7037.0 & 2455050.911367 & 0.000022 & 2 & 1 & 2 \\ 
  27 & 7167.5 & 2455065.315208 & 0.000082 & 7 & 2 & 5 \\ 
  28 & 7204.0 & 2455069.343870 & 0.000036 & 3 & 1 & 5 \\ 
  29 & 7223.0 & 2455071.440996 & 0.000014 & 1 & 1 & 5 \\ 
  30 & 7304.0 & 2455080.381278 & 0.000071 & 6 & 1 & 2 \\ 
  31 & 7322.0 & 2455082.368000 & 0.000021 & 2 & 1 & 2 \\ 
  32 & 7386.0 & 2455089.432002 & 0.000024 & 2 & 1 & 6 \\ 
  33 & 7503.0 & 2455102.345843 & 0.000018 & 2 & 1 & 5 \\ 
  34 & 7557.0 & 2455108.306027 & 0.000041 & 4 & 1 & 5 \\ 
  35 & 7955.0 & 2455152.234856 & 0.000018 & 2 & 1 & 6 \\ 
   \hline
\end{tabular}
\end{table}

\addtocounter{table}{-1}
\begin{table}
\centering
\caption{continued...} 
\begin{tabular}{crccccc}
  \hline
 & Cycle & BJD & Error[d] & Err[s] & T & Ref \\ 
  \hline
  36 & 9797.0 & 2455355.544034 & 0.000035 & 3 & 1 & 5 \\ 
  37 & 9823.5 & 2455358.468971 & 0.000046 & 4 & 2 & 2 \\ 
  38 & 9959.0 & 2455373.424740 & 0.000068 & 6 & 1 & 2 \\ 
  39 & 10131.0 & 2455392.409097 & 0.000018 & 2 & 1 & 2 \\ 
  40 & 10279.0 & 2455408.744420 & 0.000020 & 2 & 1 & 4 \\ 
  41 & 10287.0 & 2455409.627440 & 0.000020 & 2 & 1 & 4 \\ 
  42 & 10451.0 & 2455427.728770 & 0.000010 & 1 & 1 & 4 \\ 
  43 & 10593.0 & 2455443.401924 & 0.000010 & 1 & 1 & 6 \\ 
  44 & 10646.0 & 2455449.251760 & 0.000050 & 4 & 1 & 3 \\ 
  45 & 10647.0 & 2455449.362150 & 0.000020 & 2 & 1 & 3 \\ 
  46 & 10673.0 & 2455452.231890 & 0.000020 & 2 & 1 & 3 \\ 
  47 & 10918.0 & 2455479.273564 & 0.000012 & 1 & 1 & 6 \\ 
  48 & 11146.5 & 2455504.494050 & 0.000010 & 1 & 2 & 4 \\ 
  49 & 12763.0 & 2455682.913998 & 0.000013 & 1 & 1 & 2 \\ 
  50 & 12799.0 & 2455686.887451 & 0.000013 & 1 & 1 & 2 \\ 
  51 & 12799.5 & 2455686.942699 & 0.000031 & 3 & 2 & 2 \\ 
  52 & 13077.0 & 2455717.571460 & 0.000010 & 1 & 1 & 3 \\ 
  53 & 13368.0 & 2455749.690361 & 0.000006 & 1 & 1 & 4 \\ 
  54 & 13377.0 & 2455750.683717 & 0.000004 & 0 & 1 & 4 \\ 
  55 & 13469.0 & 2455760.838180 & 0.000027 & 2 & 1 & 2 \\ 
  56 & 13469.5 & 2455760.893396 & 0.000031 & 3 & 2 & 2 \\ 
  57 & 13470.0 & 2455760.948549 & 0.000011 & 1 & 1 & 2 \\ 
  58 & 13488.0 & 2455762.935315 & 0.000018 & 2 & 1 & 2 \\ 
  59 & 13511.0 & 2455765.473867 & 0.000017 & 1 & 1 & 2 \\ 
  60 & 13542.0 & 2455768.895489 & 0.000037 & 3 & 1 & 2 \\ 
  61 & 13629.0 & 2455778.498061 & 0.000009 & 1 & 1 & 4 \\ 
  62 & 13632.0 & 2455778.829154 & 0.000010 & 1 & 1 & 2 \\ 
  63 & 13682.0 & 2455784.347812 & 0.000064 & 6 & 1 & 2 \\ 
  64 & 13768.0 & 2455793.840061 & 0.000012 & 1 & 1 & 2 \\ 
  65 & 13827.0 & 2455800.352168 & 0.000014 & 1 & 1 & 2 \\ 
  66 & 13828.0 & 2455800.462510 & 0.000013 & 1 & 1 & 2 \\ 
  67 & 13845.0 & 2455802.338869 & 0.000024 & 2 & 1 & 2 \\ 
  68 & 13846.0 & 2455802.449197 & 0.000036 & 3 & 1 & 2 \\ 
  69 & 13872.0 & 2455805.318959 & 0.000015 & 1 & 1 & 2 \\ 
  70 & 13873.0 & 2455805.429322 & 0.000017 & 1 & 1 & 2 \\ 
  71 & 13899.0 & 2455808.299065 & 0.000016 & 1 & 1 & 2 \\ 
  72 & 14062.0 & 2455826.290080 & 0.000020 & 2 & 1 & 3 \\ 
  73 & 14089.0 & 2455829.270170 & 0.000020 & 2 & 1 & 3 \\ 
  74 & 14162.0 & 2455837.327516 & 0.000016 & 1 & 1 & 6 \\ 
  75 & 14379.0 & 2455861.278730 & 0.000020 & 2 & 1 & 3 \\ 
  76 & 14397.0 & 2455863.265420 & 0.000030 & 3 & 1 & 3 \\ 
  77 & 14400.0 & 2455863.596559 & 0.000010 & 1 & 1 & 2 \\ 
   \hline
\end{tabular}
\end{table}

\addtocounter{table}{-1}
\begin{table}
\centering
\caption{continued...} 
\begin{tabular}{crccccc}
  \hline
   & Cycle & BJD & Error[d] & Err[s] & T & Ref. \\
  \hline
  78 & 14406.0 & 2455864.258790 & 0.000030 & 3 & 1 & 3 \\ 
  79 & 16024.0 & 2456042.844216 & 0.000004 & 0 & 1 & 4 \\ 
  80 & 16350.0 & 2456078.826240 & 0.000008 & 1 & 1 & 4 \\ 
  81 & 16808.0 & 2456129.377577 & 0.000015 & 1 & 1 & 7 \\ 
  82 & 16808.5 & 2456129.432822 & 0.000023 & 2 & 2 & 7 \\ 
  83 & 16817.0 & 2456130.370981 & 0.000006 & 0 & 1 & 7 \\ 
  84 & 16835.0 & 2456132.357723 & 0.000017 & 1 & 1 & 6 \\ 
  85 & 16835.5 & 2456132.412907 & 0.000033 & 3 & 2 & 6 \\ 
  86 & 17019.0 & 2456152.666554 & 0.000006 & 1 & 1 & 4 \\ 
  87 & 17650.0 & 2456222.312685 & 0.000020 & 2 & 1 & 5 \\ 
  88 & 17867.0 & 2456246.263846 & 0.000021 & 2 & 1 & 5 \\ 
  89 & 19301.0 & 2456404.540320 & 0.000024 & 2 & 1 & 6 \\ 
  90 & 19545.0 & 2456431.471682 & 0.000019 & 2 & 1 & 6 \\ 
  91 & 19744.0 & 2456453.436075 & 0.000040 & 3 & 1 & 6 \\ 
  92 & 19744.5 & 2456453.491275 & 0.000099 & 9 & 2 & 6 \\ 
  93 & 20206.0 & 2456504.428991 & 0.000015 & 1 & 1 & 6 \\ 
  94 & 20206.5 & 2456504.484030 & 0.000044 & 4 & 2 & 6 \\ 
  95 & 20269.0 & 2456511.382498 & 0.000014 & 1 & 1 & 6 \\ 
  96 & 20360.0 & 2456521.426538 & 0.000013 & 1 & 1 & 6 \\ 
  97 & 20803.0 & 2456570.322299 & 0.000098 & 9 & 1 & 6 \\ 
  98 & 20812.0 & 2456571.315608 & 0.000025 & 2 & 1 & 6 \\ 
  99 & 20812.5 & 2456571.370781 & 0.000113 & 10 & 2 & 6 \\ 
  100 & 20813.0 & 2456571.425973 & 0.000046 & 4 & 1 & 6 \\ 
  101 & 20830.0 & 2456573.302321 & 0.000014 & 1 & 1 & 6 \\ 
  102 & 20975.0 & 2456589.306568 & 0.000016 & 1 & 1 & 5 \\ 
  103 & 23458.0 & 2456863.365367 & 0.000033 & 3 & 1 & 8 \\ 
  104 & 23467.0 & 2456864.358634 & 0.000032 & 3 & 1 & 8 \\ 
  105 & 24553.0 & 2456984.224844 & 0.000017 & 1 & 1 & 8 \\ 
  106 & 26132.0 & 2457158.505405 & 0.000012 & 1 & 1 & 9 \\ 
  107 & 26141.0 & 2457159.498659 & 0.000030 & 3 & 1 & 9 \\ 
  108 & 26205.0 & 2457166.562708 & 0.000012 & 1 & 1 & 8 \\ 
  109 & 26213.0 & 2457167.445649 & 0.000021 & 2 & 1 & 8 \\ 
  110 & 26331.0 & 2457180.469794 & 0.000026 & 2 & 1 & 9 \\ 
  111 & 26413.0 & 2457189.520546 & 0.000020 & 2 & 1 & 8 \\ 
  112 & 26422.0 & 2457190.513847 & 0.000010 & 1 & 1 & 8 \\ 
  113 & 26422.5 & 2457190.569067 & 0.000075 & 6 & 2 & 8 \\ 
  114 & 26648.0 & 2457215.458344 & 0.000012 & 1 & 1 & 9 \\ 
  115 & 26683.0 & 2457219.321420 & 0.000012 & 1 & 1 & 8 \\ 
  116 & 26684.0 & 2457219.431837 & 0.000013 & 1 & 1 & 8 \\ 
  117 & 26730.0 & 2457224.509018 & 0.000014 & 1 & 1 & 9 \\ 
  118 & 26846.0 & 2457237.312397 & 0.000017 & 1 & 1 & 9 \\ 
  119 & 26911.0 & 2457244.486749 & 0.000012 & 1 & 1 & 9 \\ 
   \hline
\end{tabular}
\end{table}

\addtocounter{table}{-1}
\begin{table}
\centering
\caption{continued...} 
\begin{tabular}{crccccc}
  \hline
   & Cycle & BJD & Error[d] & Err[s] & T & Ref. \\
  \hline
  120 & 26937.5 & 2457247.411658 & 0.000056 & 5 & 2 & 8 \\ 
  121 & 27009.0 & 2457255.303376 & 0.000015 & 1 & 1 & 9 \\ 
  122 & 27073.0 & 2457262.367301 & 0.000016 & 1 & 1 & 9 \\ 
  123 & 27082.0 & 2457263.360683 & 0.000034 & 3 & 1 & 9 \\ 
  124 & 27099.0 & 2457265.237010 & 0.000020 & 2 & 1 & 9 \\ 
  125 & 27154.0 & 2457271.307578 & 0.000031 & 3 & 1 & 9 \\ 
  126 & 27408.0 & 2457299.342580 & 0.000028 & 2 & 1 & 9 \\ 
  127 & 27543.0 & 2457314.243090 & 0.000010 & 1 & 1 & 9 \\ 
  128 & 28004.0 & 2457365.125453 & 0.000022 & 2 & 1 & 9 \\ 
  129 & 29411.5 & 2457520.476902 & 0.000061 & 5 & 2 & 8 \\ 
  130 & 29412.0 & 2457520.532048 & 0.000015 & 1 & 1 & 8 \\ 
  131 & 29611.0 & 2457542.496481 & 0.000012 & 1 & 1 & 9 \\ 
  132 & 29620.0 & 2457543.489845 & 0.000014 & 1 & 1 & 9 \\ 
  133 & 29647.0 & 2457546.469948 & 0.000039 & 3 & 1 & 9 \\ 
  134 & 29956.0 & 2457580.575537 & 0.000013 & 1 & 1 & 8 \\ 
  135 & 30135.0 & 2457600.332481 & 0.000014 & 1 & 1 & 9 \\ 
  136 & 30163.0 & 2457603.422952 & 0.000012 & 1 & 1 & 9 \\ 
  137 & 30172.0 & 2457604.416311 & 0.000011 & 1 & 1 & 9 \\ 
  138 & 30180.0 & 2457605.299328 & 0.000013 & 1 & 1 & 9 \\ 
  139 & 30397.0 & 2457629.250499 & 0.000009 & 1 & 1 & 8 \\ 
  140 & 30399.0 & 2457629.471256 & 0.000020 & 2 & 1 & 8 \\ 
  141 & 30408.0 & 2457630.464585 & 0.000006 & 1 & 1 & 8 \\ 
  142 & 30660.0 & 2457658.278862 & 0.000008 & 1 & 1 & 9 \\ 
  143 & 30669.0 & 2457659.272248 & 0.000008 & 1 & 1 & 9 \\ 
  144 & 30669.0 & 2457659.272263 & 0.000014 & 1 & 1 & 8 \\ 
  145 & 30670.0 & 2457659.382587 & 0.000009 & 1 & 1 & 8 \\ 
  146 & 30670.0 & 2457659.382609 & 0.000015 & 1 & 1 & 9 \\ 
  147 & 30705.0 & 2457663.245692 & 0.000008 & 1 & 1 & 9 \\ 
  148 & 30714.0 & 2457664.239053 & 0.000010 & 1 & 1 & 9 \\ 
  149 & 30904.0 & 2457685.210139 & 0.000012 & 1 & 1 & 9 \\ 
  150 & 30931.0 & 2457688.190200 & 0.000015 & 1 & 1 & 8 \\ 
  151 & 30931.0 & 2457688.190212 & 0.000011 & 1 & 1 & 9 \\ 
  152 & 30941.0 & 2457689.293987 & 0.000027 & 2 & 1 & 9 \\ 
  153 & 31004.0 & 2457696.247504 & 0.000009 & 1 & 1 & 8 \\ 
   \hline
\end{tabular}
\end{table}

\bibliographystyle{aasjournal}
\bibliography{nsvs1425}

\begin{thebibliography}{}
\expandafter\ifx\csname natexlab\endcsname\relax\def\natexlab#1{#1}\fi

\bibitem[{{Almeida} {et~al.}(2013){Almeida}, {Jablonski}, \&
  {Rodrigues}}]{Almeida2013}
{Almeida}, L.~A., {Jablonski}, F., \& {Rodrigues}, C.~V. 2013, \apj, 766, 11

\bibitem[{{Almeida} {et~al.}(2012){Almeida}, {Jablonski}, {Tello}, \&
  {Rodrigues}}]{Almeida2012}
{Almeida}, L.~A., {Jablonski}, F., {Tello}, J., \& {Rodrigues}, C.~V. 2012,
  \mnras, 423, 478

\bibitem[{{Applegate}(1992)}]{Applegate1992}
{Applegate}, J.~H. 1992, \apj, 385, 621

\bibitem[{{Bear} \& {Soker}(2014)}]{Bear2014}
{Bear}, E., \& {Soker}, N. 2014, \mnras, 444, 1698

\bibitem[{{Beuermann} {et~al.}(2012{\natexlab{a}}){Beuermann}, {Dreizler},
  {Hessman}, \& {Deller}}]{Beuermann2012b}
{Beuermann}, K., {Dreizler}, S., {Hessman}, F.~V., \& {Deller}, J.
  2012{\natexlab{a}}, \aap, 543, A138

\bibitem[{{Beuermann} {et~al.}(2010){Beuermann}, {Hessman}, {Dreizler},
  {Marsh}, {Parsons}, {Winget}, {Miller}, {Schreiber}, {Kley}, {Dhillon},
  {Littlefair}, {Copperwheat}, \& {Hermes}}]{Beuermann2010}
{Beuermann}, K., {Hessman}, F.~V., {Dreizler}, S., {et~al.} 2010, \aap, 521,
  L60

\bibitem[{{Beuermann} {et~al.}(2011){Beuermann}, {Buhlmann}, {Diese},
  {Dreizler}, {Hessman}, {Husser}, {Miller}, {Nickol}, {Pons}, {Ruhr},
  {Schm{\"u}lling}, {Schwope}, {Sorge}, {Ulrichs}, {Winget}, \&
  {Winget}}]{Beuermann2011}
{Beuermann}, K., {Buhlmann}, J., {Diese}, J., {et~al.} 2011, \aap, 526, A53

\bibitem[{{Beuermann} {et~al.}(2012{\natexlab{b}}){Beuermann}, {Breitenstein},
  {Debski}, {Diese}, {Dubovsky}, {Dreizler}, {Hessman}, {Hornoch}, {Husser},
  {Pojmanski}, {Wolf}, {Wo{\'z}niak}, {Zasche}, {Denk}, {Langer}, {Wagner},
  {Wahrenberg}, {Bollmann}, {Habermann}, {Haustovich}, {Lauser}, {Liebing}, \&
  {Niederstadt}}]{Beuermann2012}
{Beuermann}, K., {Breitenstein}, P., {Debski}, B., {et~al.} 2012{\natexlab{b}},
  \aap, 540, A8

\bibitem[{{Bours} {et~al.}(2014){Bours}, {Marsh}, {Breedt}, {Copperwheat},
  {Dhillon}, {Leckngam}, {Littlefair}, {Parsons}, \& {Prasit}}]{Bours2014}
{Bours}, M.~C.~P., {Marsh}, T.~R., {Breedt}, E., {et~al.} 2014, \mnras, 445,
  1924

\bibitem[{{Brinkworth} {et~al.}(2006){Brinkworth}, {Marsh}, {Dhillon}, \&
  {Knigge}}]{Brinkworth2006}
{Brinkworth}, C.~S., {Marsh}, T.~R., {Dhillon}, V.~S., \& {Knigge}, C. 2006,
  \mnras, 365, 287

\bibitem[{{Charbonneau}(1995)}]{Charbonneau1995}
{Charbonneau}, P. 1995, \apjs, 101, 309

\bibitem[{{Deeg} {et~al.}(2008){Deeg}, {Oca{\~n}a}, {Kozhevnikov},
  {Charbonneau}, {O'Donovan}, \& {Doyle}}]{Deeg2008}
{Deeg}, H.~J., {Oca{\~n}a}, B., {Kozhevnikov}, V.~P., {et~al.} 2008, \aap, 480,
  563

\bibitem[{{Doyle} {et~al.}(2011){Doyle}, {Carter}, {Fabrycky}, {Slawson},
  {Howell}, {Winn}, {Orosz}, {Prsa}, {Welsh}, {Quinn}, {Latham}, {Torres},
  {Buchhave}, {Marcy}, {Fortney}, {Shporer}, {Ford}, {Lissauer}, {Ragozzine},
  {Rucker}, {Batalha}, {Jenkins}, {Borucki}, {Koch}, {Middour}, {Hall},
  {McCauliff}, {Fanelli}, {Quintana}, {Holman}, {Caldwell}, {Still},
  {Stefanik}, {Brown}, {Esquerdo}, {Tang}, {Furesz}, {Geary}, {Berlind},
  {Calkins}, {Short}, {Steffen}, {Sasselov}, {Dunham}, {Cochran}, {Boss},
  {Haas}, {Buzasi}, \& {Fischer}}]{Doyle2011}
{Doyle}, L.~R., {Carter}, J.~A., {Fabrycky}, D.~C., {et~al.} 2011, Science,
  333, 1602

\bibitem[{{Eastman} {et~al.}(2010){Eastman}, {Siverd}, \&
  {Gaudi}}]{Eastman2010}
{Eastman}, J., {Siverd}, R., \& {Gaudi}, B.~S. 2010, \pasp, 122, 935

\bibitem[{{Foreman-Mackey} {et~al.}(2013){Foreman-Mackey}, {Hogg}, {Lang}, \&
  {Goodman}}]{Foreman2013}
{Foreman-Mackey}, D., {Hogg}, D.~W., {Lang}, D., \& {Goodman}, J. 2013, \pasp,
  125, 306

\bibitem[{{Goodman} \& {Weare}(2010)}]{Goodman2010}
{Goodman}, J., \& {Weare}, J. 2010, Comm. Apl. Math and Comp. Sci., 1, 65

\bibitem[{{Go{\'z}dziewski} {et~al.}(2012){Go{\'z}dziewski}, {Nasiroglu},
  {S{\l}owikowska}, {Beuermann}, {Kanbach}, {Gauza}, {Maciejewski}, {Schwarz},
  {Schwope}, {Hinse}, {Haghighipour}, {Burwitz}, {S{\l}onina}, \&
  {Rau}}]{Gozdziewski2012}
{Go{\'z}dziewski}, K., {Nasiroglu}, I., {S{\l}owikowska}, A., {et~al.} 2012,
  \mnras, 425, 930

\bibitem[{{Go{\'z}dziewski} {et~al.}(2015){Go{\'z}dziewski}, {S{\l}owikowska},
  {Dimitrov}, {Krzeszowski}, {{\.Z}ejmo}, {Kanbach}, {Burwitz}, {Rau},
  {Irawati}, {Richichi}, {Gawro{\'n}ski}, {Nowak}, {Nasiroglu}, \&
  {Kubicki}}]{Gozdziewski2015}
{Go{\'z}dziewski}, K., {S{\l}owikowska}, A., {Dimitrov}, D., {et~al.} 2015,
  \mnras, 448, 1118

\bibitem[{{Hardy} {et~al.}(2015){Hardy}, {Schreiber}, {Parsons}, {Caceres},
  {Retamales}, {Wahhaj}, {Mawet}, {Canovas}, {Cieza}, {Marsh}, {Bours},
  {Dhillon}, \& {Bayo}}]{Hardy2015}
{Hardy}, A., {Schreiber}, M.~R., {Parsons}, S.~G., {et~al.} 2015, \apjl, 800,
  L24

\bibitem[{{Hinse} {et~al.}(2012){Hinse}, {Lee}, {Go{\'z}dziewski},
  {Haghighipour}, {Lee}, \& {Scullion}}]{Hinse2012}
{Hinse}, T.~C., {Lee}, J.~W., {Go{\'z}dziewski}, K., {et~al.} 2012, \mnras,
  420, 3609

\bibitem[{{Hinse} {et~al.}(2014){Hinse}, {Lee}, {Go{\'z}dziewski}, {Horner}, \&
  {Wittenmyer}}]{Hinse2014}
{Hinse}, T.~C., {Lee}, J.~W., {Go{\'z}dziewski}, K., {Horner}, J., \&
  {Wittenmyer}, R.~A. 2014, \mnras, 438, 307

\bibitem[{{Holman} \& {Wiegert}(1999)}]{Holman1999}
{Holman}, M.~J., \& {Wiegert}, P.~A. 1999, \aj, 117, 621

\bibitem[{{Horner} {et~al.}(2012){Horner}, {Hinse}, {Wittenmyer}, {Marshall},
  \& {Tinney}}]{Horner2012}
{Horner}, J., {Hinse}, T.~C., {Wittenmyer}, R.~A., {Marshall}, J.~P., \&
  {Tinney}, C.~G. 2012, \mnras, 427, 2812

\bibitem[{{Horner} {et~al.}(2013){Horner}, {Wittenmyer}, {Hinse}, {Marshall},
  {Mustill}, \& {Tinney}}]{Horner2013}
{Horner}, J., {Wittenmyer}, R.~A., {Hinse}, T.~C., {et~al.} 2013, \mnras, 435,
  2033

\bibitem[{{Irwin}(1952)}]{Irwin1952}
{Irwin}, J.~B. 1952, \apj, 116, 211

\bibitem[{{Kilkenny} \& {Koen}(2012)}]{Kilkenny2012}
{Kilkenny}, D., \& {Koen}, C. 2012, \mnras, 421, 3238

\bibitem[{{Kostov} {et~al.}(2016{\natexlab{a}}){Kostov}, {Moore}, {Tamayo},
  {Jayawardhana}, \& {Rinehart}}]{Kostov2016b}
{Kostov}, V.~B., {Moore}, K., {Tamayo}, D., {Jayawardhana}, R., \& {Rinehart},
  S.~A. 2016{\natexlab{a}}, \apj, 832, 183

\bibitem[{{Kostov} {et~al.}(2016{\natexlab{b}}){Kostov}, {Orosz}, {Welsh},
  {Doyle}, {Fabrycky}, {Haghighipour}, {Quarles}, {Short}, {Cochran}, {Endl},
  {Ford}, {Gregorio}, {Hinse}, {Isaacson}, {Jenkins}, {Jensen}, {Kane}, {Kull},
  {Latham}, {Lissauer}, {Marcy}, {Mazeh}, {M{\"u}ller}, {Pepper}, {Quinn},
  {Ragozzine}, {Shporer}, {Steffen}, {Torres}, {Windmiller}, \&
  {Borucki}}]{Kostov2016a}
{Kostov}, V.~B., {Orosz}, J.~A., {Welsh}, W.~F., {et~al.} 2016{\natexlab{b}},
  \apj, 827, 86

\bibitem[{{Lanza}(2006)}]{Lanza2006}
{Lanza}, A.~F. 2006, \mnras, 369, 1773

\bibitem[{{Lanza} \& {Rodon{\`o}}(1999)}]{Lanza1999}
{Lanza}, A.~F., \& {Rodon{\`o}}, M. 1999, \aap, 349, 887

\bibitem[{{Lanza} {et~al.}(1998){Lanza}, {Rodono}, \& {Rosner}}]{Lanza1998}
{Lanza}, A.~F., {Rodono}, M., \& {Rosner}, R. 1998, \mnras, 296, 893

\bibitem[{{Lee} {et~al.}(2014){Lee}, {Hinse}, {Youn}, \& {Han}}]{Lee2014}
{Lee}, J.~W., {Hinse}, T.~C., {Youn}, J.-H., \& {Han}, W. 2014, \mnras, 445,
  2331

\bibitem[{{Lee} {et~al.}(2009){Lee}, {Kim}, {Kim}, {Koch}, {Lee}, {Kim}, \&
  {Park}}]{Lee2009}
{Lee}, J.~W., {Kim}, S.-L., {Kim}, C.-H., {et~al.} 2009, \aj, 137, 3181

\bibitem[{{Lohr} {et~al.}(2014){Lohr}, {Norton}, {Anderson}, {Collier Cameron},
  {Faedi}, {Haswell}, {Hellier}, {Hodgkin}, {Horne}, {Kolb}, {Maxted},
  {Pollacco}, {Skillen}, {Smalley}, {West}, \& {Wheatley}}]{Lohr2014}
{Lohr}, M.~E., {Norton}, A.~J., {Anderson}, D.~R., {et~al.} 2014, \aap, 566,
  A128

\bibitem[{{Marsh} {et~al.}(2014){Marsh}, {Parsons}, {Bours}, {Littlefair},
  {Copperwheat}, {Dhillon}, {Breedt}, {Caceres}, \& {Schreiber}}]{Marsh2014}
{Marsh}, T.~R., {Parsons}, S.~G., {Bours}, M.~C.~P., {et~al.} 2014, \mnras,
  437, 475

\bibitem[{{Pojmanski}(1997)}]{Pojmanski1997}
{Pojmanski}, G. 1997, \actaa, 47, 467

\bibitem[{{Portegies Zwart}(2013)}]{PortegiesZwart2013}
{Portegies Zwart}, S. 2013, \mnras, 429, L45

\bibitem[{{Potter} {et~al.}(2011){Potter}, {Romero-Colmenero}, {Ramsay},
  {Crawford}, {Gulbis}, {Barway}, {Zietsman}, {Kotze}, {Buckley}, {O'Donoghue},
  {Siegmund}, {McPhate}, {Welsh}, \& {Vallerga}}]{Potter2011}
{Potter}, S.~B., {Romero-Colmenero}, E., {Ramsay}, G., {et~al.} 2011, \mnras,
  416, 2202

\bibitem[{{Pribulla } {et~al.}(2012){Pribulla }, {Va{\v n}ko}, {Ammler-von
  Eiff}, {Andreev}, {Aslant{\"u}rk}, {Awadalla}, {Balu{\v d}ansk{\'y}},
  {Bonanno}, {Bo{\v z}i{\'c}}, {Catanzaro}, {{\c C}elik}, {Christopoulou},
  {Covino}, {Cusano}, {Dimitrov}, {Dubovsk{\'y}}, {Eigmueller}, {Esmer},
  {Frasca}, {Hamb{\'a}lek}, {Hanna}, {Hanslmeier}, {Kalomeni}, {Kjurkchieva},
  {Krushevska}, {Kudzej}, {Kundra}, {Kuznyetsova}, {Lee}, {Leitzinger},
  {Maciejewski}, {Moldovan}, {Morais}, {Mugrauer}, {Neuh{\"a}user},
  {Niedzielski}, {Odert}, {Ohlert}, {{\"O}zavc{\i}}, {Papageorgiou},
  {Parimucha}, {Poddan{\'y}}, {Pop}, {Raetz}, {Raetz}, {Romanyuk}, {Ru{\v
  z}djak}, {Schulz}, {{\c S}enavc{\i}}, {Srdoc}, {Szalai}, {Sz{\'e}kely},
  {Sudar}, {Tezcan}, {T{\"o}r{\"u}n}, {Turcu}, {Vince}, \&
  {Zejda}}]{Pribulla2012}
{Pribulla }, T., {Va{\v n}ko}, M., {Ammler-von Eiff}, M., {et~al.} 2012,
  Astronomische Nachrichten, 333, 754

\bibitem[{{Qian} {et~al.}(2012{\natexlab{a}}){Qian}, {Liu}, {Zhu}, {Dai},
  {Fern{\'a}ndez Laj{\'u}s}, \& {Baume}}]{Qian2012a}
{Qian}, S.-B., {Liu}, L., {Zhu}, L.-Y., {et~al.} 2012{\natexlab{a}}, \mnras,
  422, 24

\bibitem[{{Qian} {et~al.}(2012{\natexlab{b}}){Qian}, {Zhu}, {Dai},
  {Fern{\'a}ndez-Laj{\'u}s}, {Xiang}, \& {He}}]{Qian2012b}
{Qian}, S.-B., {Zhu}, L.-Y., {Dai}, Z.-B., {et~al.} 2012{\natexlab{b}}, \apjl,
  745, L23

\bibitem[{{Qian} {et~al.}(2010){Qian}, {Zhu}, {Liu}, {Dai}, {He}, {Liao},
  {Zhao}, {Zola}, {Kuligowska}, \& {Winiarski}}]{Qian2010}
{Qian}, S.-B., {Zhu}, L.-Y., {Liu}, L., {et~al.} 2010, \apss, 329, 113

\bibitem[{{Qian} {et~al.}(2011){Qian}, {Liu}, {Liao}, {Li}, {Zhu}, {Dai}, {He},
  {Zhao}, {Zhang}, \& {Li}}]{Qian2011}
{Qian}, S.-B., {Liu}, L., {Liao}, W.-P., {et~al.} 2011, \mnras, 414, L16

\bibitem[{{Qian} {et~al.}(2013){Qian}, {Shi}, {Zola}, {Koziel-Wierzbowska},
  {Winiarski}, {Szymanski}, {Ogloza}, {Li}, {Zhu}, {Liu}, {He}, {Liao}, {Zhao},
  {Wang}, {Zhang}, \& {Jiang}}]{Qian2013}
{Qian}, S.-B., {Shi}, G., {Zola}, S., {et~al.} 2013, \mnras, 436, 1408

\bibitem[{{Ribas}(2005)}]{Ribas2005}
{Ribas}, I. 2005, in Astronomical Society of the Pacific Conference Series,
  Vol. 335, The Light-Time Effect in Astrophysics: Causes and cures of the O-C
  diagram, ed. C.~{Sterken}, 55

\bibitem[{{Schwarz} {et~al.}(2009){Schwarz}, {Schwope}, {Vogel}, {Dhillon},
  {Marsh}, {Copperwheat}, {Littlefair}, \& {Kanbach}}]{Schwarz2009}
{Schwarz}, R., {Schwope}, A.~D., {Vogel}, J., {et~al.} 2009, \aap, 496, 833

\bibitem[{{Schwope} \& {Thinius}(2014)}]{Schwope2014}
{Schwope}, A.~D., \& {Thinius}, B.~D. 2014, Astronomische Nachrichten, 335, 357

\bibitem[{{Schwope} {et~al.}(1993){Schwope}, {Thomas}, \&
  {Beuermann}}]{Schwope1993}
{Schwope}, A.~D., {Thomas}, H.~C., \& {Beuermann}, K. 1993, \aap, 271, L25

\bibitem[{{Tian} {et~al.}(2009){Tian}, {Xiang}, \& {Tao}}]{Tian2009}
{Tian}, Y.~P., {Xiang}, F.~Y., \& {Tao}, X. 2009, \apss, 319, 119

\bibitem[{{Veras} {et~al.}(2017){Veras}, {Georgakarakos}, {Dobbs-Dixon}, \&
  {G{\"a}nsicke}}]{Veras2017}
{Veras}, D., {Georgakarakos}, N., {Dobbs-Dixon}, I., \& {G{\"a}nsicke}, B.~T.
  2017, \mnras, 465, 2053

\bibitem[{{Veras} \& {Tout}(2012)}]{Veras2012}
{Veras}, D., \& {Tout}, C.~A. 2012, \mnras, 422, 1648

\bibitem[{{Veras} {et~al.}(2011){Veras}, {Wyatt}, {Mustill}, {Bonsor}, \&
  {Eldridge}}]{Veras2011}
{Veras}, D., {Wyatt}, M.~C., {Mustill}, A.~J., {Bonsor}, A., \& {Eldridge},
  J.~J. 2011, \mnras, 417, 2104

\bibitem[{{V{\"o}lschow} {et~al.}(2014){V{\"o}lschow}, {Banerjee}, \&
  {Hessman}}]{Volschow2014}
{V{\"o}lschow}, M., {Banerjee}, R., \& {Hessman}, F.~V. 2014, \aap, 562, A19

\bibitem[{{V{\"o}lschow} {et~al.}(2016){V{\"o}lschow}, {Schleicher},
  {Perdelwitz}, \& {Banerjee}}]{Volschow2016}
{V{\"o}lschow}, M., {Schleicher}, D.~R.~G., {Perdelwitz}, V., \& {Banerjee}, R.
  2016, \aap, 587, A34

\bibitem[{{Wils} {et~al.}(2007){Wils}, {di Scala}, \& {Otero}}]{Wils2007}
{Wils}, P., {di Scala}, G., \& {Otero}, S.~A. 2007, Information Bulletin on
  Variable Stars, 5800

\bibitem[{{Wittenmyer} {et~al.}(2013){Wittenmyer}, {Horner}, \&
  {Marshall}}]{Wittenmyer2013}
{Wittenmyer}, R.~A., {Horner}, J., \& {Marshall}, J.~P. 2013, \mnras, 431, 2150

\bibitem[{{Wo{\'z}niak} {et~al.}(2004{\natexlab{a}}){Wo{\'z}niak}, {Williams},
  {Vestrand}, \& {Gupta}}]{Wozniak2004}
{Wo{\'z}niak}, P.~R., {Williams}, S.~J., {Vestrand}, W.~T., \& {Gupta}, V.
  2004{\natexlab{a}}, \aj, 128, 2965

\bibitem[{{Wo{\'z}niak} {et~al.}(2004{\natexlab{b}}){Wo{\'z}niak}, {Vestrand},
  {Akerlof}, {Balsano}, {Bloch}, {Casperson}, {Fletcher}, {Gisler}, {Kehoe},
  {Kinemuchi}, {Lee}, {Marshall}, {McGowan}, {McKay}, {Rykoff}, {Smith},
  {Szymanski}, \& {Wren}}]{Wozniak2004nsvs}
{Wo{\'z}niak}, P.~R., {Vestrand}, W.~T., {Akerlof}, C.~W., {et~al.}
  2004{\natexlab{b}}, \aj, 127, 2436

\bibitem[{{Zhu} {et~al.}(2011){Zhu}, {Qian}, {Liu}, {Liao}, {He}, {Li}, {Zhao},
  {Dai}, {Zhang}, \& {Li}}]{Zhu2011}
{Zhu}, L., {Qian}, S., {Liu}, L., {et~al.} 2011, in Astronomical Society of the
  Pacific Conference Series, Vol. 451, 9th Pacific Rim Conference on Stellar
  Astrophysics, ed. S.~{Qain}, K.~{Leung}, L.~{Zhu}, \& S.~{Kwok}, 155

\bibitem[{{Zorotovic} \& {Schreiber}(2013)}]{Zorotovic2013}
{Zorotovic}, M., \& {Schreiber}, M.~R. 2013, \aap, 549, A95

\end{thebibliography}

\listofchanges

\end{document}